\documentclass[11pt,paper=a4,DIV=12,abstract=true]{scrartcl}

\usepackage[T1]{fontenc}
\usepackage[utf8]{inputenc}
\usepackage{lmodern}
\usepackage{microtype}

\usepackage{amsmath,amssymb,amsthm,mathtools}
\usepackage{mathrsfs}
\usepackage{bm}
\usepackage[shortlabels]{enumitem}

\usepackage{graphicx}
\usepackage{booktabs}
\usepackage{threeparttable}

\usepackage{aliascnt}
\usepackage[dvipsnames]{xcolor}

\usepackage{algorithm}
\usepackage{algorithmic}

\usepackage[round,authoryear]{natbib}
\usepackage[unicode=true]{hyperref}
\usepackage[nameinlink,noabbrev,capitalize]{cleveref}
\setkomafont{disposition}{\normalfont\bfseries}
\addtokomafont{section}{\large}
\addtokomafont{subsection}{\normalsize}
\setkomafont{title}{\normalfont\bfseries}
\setkomafont{author}{\normalfont}
\setkomafont{date}{\normalfont}
\setkomafont{caption}{\small}
\setkomafont{captionlabel}{\small\bfseries}

\numberwithin{equation}{section}

\theoremstyle{plain}
\newtheorem{theorem}{Theorem}[section]

\newaliascnt{proposition}{theorem}

\aliascntresetthe{proposition}

\newaliascnt{lemma}{theorem}
\newtheorem{lemma}[lemma]{Lemma}
\aliascntresetthe{lemma}

\newaliascnt{corollary}{theorem}
\newtheorem{corollary}[corollary]{Corollary}
\aliascntresetthe{corollary}

\theoremstyle{definition}

\newaliascnt{assumption}{theorem}

\aliascntresetthe{assumption}

\newaliascnt{definition}{theorem}

\aliascntresetthe{definition}

\newaliascnt{example}{theorem}
\newtheorem{example}[example]{Example}
\aliascntresetthe{example}

\theoremstyle{remark}

\newaliascnt{remark}{theorem}
\newtheorem{remark}[remark]{Remark}
\aliascntresetthe{remark}

\crefname{theorem}{Theorem}{Theorems}
\Crefname{theorem}{Theorem}{Theorems}
\crefname{proposition}{Proposition}{Propositions}
\Crefname{proposition}{Proposition}{Propositions}
\crefname{lemma}{Lemma}{Lemmas}
\Crefname{lemma}{Lemma}{Lemmas}
\crefname{corollary}{Corollary}{Corollaries}
\Crefname{corollary}{Corollary}{Corollaries}
\crefname{assumption}{Assumption}{Assumptions}
\Crefname{assumption}{Assumption}{Assumptions}
\crefname{definition}{Definition}{Definitions}
\Crefname{definition}{Definition}{Definitions}
\crefname{example}{Example}{Examples}
\Crefname{example}{Example}{Examples}
\crefname{remark}{Remark}{Remarks}
\Crefname{remark}{Remark}{Remarks}

\DeclareMathOperator{\Var}{Var}
\DeclareMathOperator{\Cov}{Cov}

\newcommand{\R}{\mathbb{R}}
\newcommand{\E}{\mathbb{E}}
\newcommand{\Prob}{\mathbb{P}}

\DeclarePairedDelimiter{\abs}{\lvert}{\rvert}
\DeclarePairedDelimiter{\norm}{\lVert}{\rVert}

\makeatletter
\let\old@fnsymbol\@fnsymbol
\renewcommand{\@fnsymbol}[1]{%
  \ifnum#1=2
    \TextOrMath{\textasteriskcentered\textasteriskcentered}{**}%
  \else
    \old@fnsymbol{#1}%
  \fi
}
\makeatother

\allowdisplaybreaks
\newcommand{\keywords}[1]{\par\medskip\noindent\textbf{Keywords.} #1}
\newcommand{\msc}[1]{\par\smallskip\noindent\textbf{MSC 2020.} #1}
\newcommand{\jel}[1]{\par\smallskip\noindent\textbf{JEL classification.} #1}

\hypersetup{
  hidelinks,
  pdftitle={On the Continuity of Conditional Distributions and Stratification for Derivative Pricing},
  pdfauthor={Aleksandar Arandjelović and Patrick Cheridito},
  pdfkeywords={Stratified Monte Carlo; variance reduction; conditional expectations; conditional distributions; neural networks; derivative pricing}
}

\title{{\Large On the Continuity of Conditional Distributions\\ and Stratification for Derivative Pricing}}

\author{%
Aleksandar Arandjelović\thanks{Department of Mathematics, ETH Zürich, Zürich, Switzerland, and Institute for Statistics and Mathematics, Vienna University of Economics and Business, Vienna, Austria. Email: \href{mailto:aleksandar.arandjelovic@wu.ac.at}{aleksandar.arandjelovic@wu.ac.at}. Corresponding author.}
\and
Patrick Cheridito\thanks{Department of Mathematics, ETH Zürich, Zürich, Switzerland.}
}

\date{October 23, 2026}

\begin{document}

\maketitle

\begin{abstract}
Stratification can reduce the estimator variance in Monte Carlo simulations by partitioning the sample space according to a low-dimensional factor and sampling separately within the resulting strata, making it well suited to the valuation of path-dependent claims in high-dimensional financial models.
Its performance, however, depends critically on the choice of the stratification factor, and effective low-dimensional factors are generally difficult to identify in complex continuous-time models.
Neural networks provide a flexible approximation class for this purpose and have proved useful in related variance-reduction problems in quantitative finance.
In this paper, we study the construction of neural stratification factors for derivative pricing under proportional allocation in continuous-time models.
Since the numerical construction of a stratification factor typically involves approximation, we establish continuity results for the associated conditional expectations and, more generally, conditional distributions, and provide several counterexamples showing the limits of this continuity theory.
Combined with density results for neural networks, our findings yield a universal approximation theorem for admissible stratification factors.
We conclude with numerical experiments involving a constant-maturity-swap range accrual in a two-factor Gaussian short-rate model and a worst-of autocallable in a Wishart stochastic covariance model, where the learned factors outperform all considered benchmarks.
\end{abstract}

\msc{Primary 91G60; Secondary 60B10, 68T07}
\jel{C15; C45; G13}

\clearpage

\section{Introduction}\label{sec:introduction}
Many valuation and risk measurement problems in quantitative finance involve expectations that are not available in closed form.
This often occurs when a model contains several risk factors or a payoff depends on the path of the underlying process.
Let $X$ be a possibly multivariate stochastic process under a probability measure $\Prob$, interpreted where appropriate as a pricing measure, and let $H(X)$ be an integrable, possibly path-dependent discounted payoff.
The quantity of interest is then
\begin{equation*}
V=\E[H(X)].
\end{equation*}
Monte Carlo methods estimate $V$ by averaging $H(X)$ over simulated realisations of $X$.
This approach extends naturally to models with intricate dependence structures and high-dimensional state spaces, provided that $X$ can be simulated and $H(X)$ evaluated pathwise.
Continuing advances in computing power further expand the scale and complexity of the financial problems that can be treated in this way.

The use of random sampling as a numerical method predates its application in finance.
An early systematic account was given by \citet{metropolis49montecarlo}, who presented Monte Carlo as a statistical approach to problems in mathematical physics.
\citet{boyle77options} subsequently developed a simulation method for option valuation under risk-neutral pricing.
The scope of financial simulation was extended by \citet{kemna90pricing} to options on average asset values and by \citet{longstaff01valuing} to American options in path-dependent and multifactor settings.
\citet{broadie06exact} later developed exact simulation methods for stochastic-volatility and affine jump-diffusion models.
Together, these contributions established Monte Carlo simulation as a general framework for valuing financial claims with increasingly complex contractual and model features.

For independent copies $X^{(1)},\ldots,X^{(N)}$ of $X$, let $\widehat V_N\coloneqq N^{-1}\sum_{i=1}^N H(X^{(i)})$ denote the crude Monte Carlo estimator.
If $v\coloneqq \Var(H(X))<\infty$, then $\Var(\widehat V_N)=v/N$, so $\widehat V_N$ converges in $L^2$ at the rate of $N^{-1/2}$.
Crucially, this rate is independent of the dimension of $X$, which is an advantage over deterministic methods such as finite-difference or quadrature schemes, whose cost typically grows exponentially with the dimension.
However, a standard confidence interval for $V$ has width proportional to $\sqrt{v/N}$, so halving it requires an increase in the number of paths by a factor of four.
The problem is particularly acute when the payoff variance $v$ is large or when high relative accuracy is required for a small expected payoff.
Deep out-of-the-money calls illustrate the latter case: $\Prob(H(X)\neq 0)$ is small, so most simulated payoffs are zero, and the relative variance $\Var(H(X))/V^2$ can be large even when $v$ is small.
Both situations can make crude Monte Carlo expensive, especially when simulating paths of $X$ is costly.

This variance problem has led to several methods for improving simulation accuracy without increasing $N$, including antithetic variates, control variates, importance sampling, stratified sampling, and quasi-Monte Carlo; see, e.g., \citet[Chapters~4--5]{glasserman03montecarlo}.
Control variates use correlated quantities with known expectations, while importance sampling changes the sampling measure to place more weight on regions that materially affect the target expectation (e.g., by increasing the probability of the event $\{H(X)\neq0\}$).
\citet{hinds23neural} use neural networks to learn control variates for stochastic differential equations.
For importance sampling, \citet{glasserman99asymptotically} derive asymptotically optimal changes of measure for path-dependent options, while \citet{guasoni08optimal} obtain explicit continuous-time drift changes in the Black--Scholes model, and \citet{lemaire10unconstrained} develop recursive schemes with convergence guarantees.
The deep importance sampling approach of \citet{arandjelovic25importance} uses neural networks to learn changes of measure for complex path-dependent models.

We now turn to stratified sampling, which is closely related to importance sampling in spirit.
Given a possibly vector-valued factor $Y$ that is not independent from $X$, it partitions the range of $Y$ into disjoint strata, samples $H(X)$ conditionally on $Y$ lying in each stratum, and estimates $V$ by a probability-weighted average of the resulting stratum means.
For square-integrable $H(X)$, the law of total variance indicates that a natural limiting criterion for choosing the stratification factor $Y$ under proportional allocation is to maximise $\Var(\E[H(X)\mid\sigma(Y)])$.
Combining large-deviation theory with Taylor approximation, \citet{glasserman99asymptotically} study optimal stratification for path-dependent options in high-dimensional, discrete-time Gaussian models.
Subsequent work studies adaptive allocation \citep{etore10adaptive}, adaptive stratification \citep{etore11adaptive}, several stratification directions \citep{jourdain11multiple}, and path-space strata based on functional quantisation \citep{corlay15functional}.

For square-integrable $H(X)$, we keep the dependence on $X$ and $H$ implicit and write
\begin{equation}
m_Y\coloneqq \E[H(X)\mid\sigma(Y)], \qquad \mu_Y\coloneqq \mathcal L\bigl(H(X)\mid\sigma(Y)\bigr),
\end{equation}
for the associated conditional expectation and conditional distribution, respectively.
In practice, $Y$ is often replaced by a learned approximation, so the criterion $\Var(\E[H(X)\mid\sigma(Y)])$ is useful in numerical applications when it remains stable under the approximation of the stratification factor.
This provides a natural motivation for studying continuity of the maps $Y\mapsto m_Y$ and $Y\mapsto \mu_Y$ with respect to convergence of the stratification factor.
Such stability would ensure that replacing an optimal stratification factor by a numerical approximation does not change the optimisation criterion, or the conditional distribution of the payoff, discontinuously.

Earlier work has studied this problem from several directions.
\citet{bryc92stability} and \citet{millet95small} examine perturbations of the conditioning variable, the latter for Gaussian factors, whereas \citet{fetter77continuity} and \citet{alonso98lp} formulate continuity through convergence of the generated $\sigma$-fields.
Other work addresses the Hausdorff metric on $\sigma$-fields \citep{boylan71equiconvergence,neveu72tightness}, strong convergence of $\sigma$-fields \citep{kudo74strong}, and weak convergence of filtrations \citep{coquet01weak}.
Our result gives a sufficient condition in terms of convergence of deterministic integrands and eventual equality of covariance ranks, without requiring equality of the lower and upper limits of the generated $\sigma$-fields.
For conditional distributions, \citet{sweeting89conditional} develops notions of conditional weak convergence, and \citet{lumpp25weak} show that, even for Gaussian vectors, convergence of the joint distribution does not necessarily imply convergence of the conditional distribution near a singular limit.
These results concern the behaviour of conditional distributions when the conditioning variable, and possibly the generated $\sigma$-field, changes.
By contrast, \citet{zabell79continuous}, \citet{lagatta13continuous}, and \citet{dolera23lipschitz} study continuity in the observed value $y$ of a kernel $K(y,\cdot)$ representing the regular conditional distribution of $H(X)$ given $Y$, so that, in the notation above, $\mu_Y=K(Y,\cdot)$ and $m_Y = \int z\, K(Y,\mathrm{d}z)$ almost surely.
We also note that convergence of $m_{Y_n}$ to $m_Y$ need not imply convergence of $\mu_{Y_n}$ to $\mu_Y$.
To see this, let $G$ and $W$ be independent standard normal random variables, set $U=\mathbf 1_{\{G\leq 0\}}+2\mathbf 1_{\{G>0\}}$, $X=UW$, and $H(x)=x$, and define $Y_n=G/n$ and $Y=0$.
Then $Y_n\to Y$ in $L^2$, and $m_{Y_n}=m_Y=0$ for every $n$, whereas $\mu_{Y_n}=\mathcal N(0,U^2)$ and $\mu_Y=\frac12\mathcal N(0,1)+\frac12\mathcal N(0,4)$.

The continuity results discussed above are not constructive: they establish stability but do not indicate how a suitable stratification factor should be chosen in practice.
This limitation is increasingly relevant as advances in computing power make it feasible to simulate models with many risk factors, path dependence, and nonlinear contractual features, settings in which explicit stratification directions of the type derived by \citet{glasserman99asymptotically} need not be available.
Choosing a useful stratification factor is then itself a nontrivial numerical problem.
Neural networks, already used for control variates and importance sampling in quantitative finance \citep{hinds23neural,arandjelovic25importance}, are well suited to this task: feedforward networks can approximate large classes of functions \citep{leshno93multilayer}, and stochastic-gradient methods allow their parameters to be trained from simulated data \citep{ghadimi13stochastic}.
This motivates writing $Y_\theta$ for a possibly vector-valued stratification factor determined by a neural network with parameter $\theta$, and choosing $\theta$ so as to maximise the variance of $m_{Y_\theta}$.
By contrast, recent approaches instead learn the strata directly from simulation output \citep{chopin25adaptive,ban26lestrat} rather than learning a stratification factor.

Motivated by the difficulty of identifying effective stratification factors in complex financial models and by the successful use of deep learning for variance reduction in quantitative finance, we study optimal stratification for the Monte Carlo valuation of financial derivatives through neural stratification factors.
The main contributions of this work are threefold:
\begin{enumerate}[label=\Roman*.]
\item We formulate optimal stratification as a stochastic optimisation problem over the parameters of a neural stratification factor in a general continuous-time setting that subsumes many financial models and payoffs, including multidimensional Itô diffusions and path-dependent options in multifactor stochastic-volatility models.
Concretely, when $X$ is driven by a continuous local martingale $M$ with deterministic covariation, we consider factors $\mathcal I(f) \coloneqq \int_0^1 f(t)^\top \mathrm{d}M_t$ for $f \in \Lambda^2$, where $\Lambda^2$ is the Hilbert space of deterministic integrands associated with the covariation measure of $M$.
\item We establish continuity of the conditional expectation $m_Y$ and conditional distribution $\mu_Y$ along sequences of Gaussian stratification factors converging in $L^2$ to a non-degenerate Gaussian limit, under mild assumptions on the driving local martingale $M$, the distribution of $X$ and the payoff $H(X)$.
As a direct corollary, we obtain a universal approximation result, showing that any admissible non-degenerate stratification factor $Y$ can be approximated, for any error tolerance $\varepsilon > 0$, by a neural stratification factor $Y_\theta$ with $\Var(m_{Y_\theta})\geq \Var(m_Y)-\varepsilon$.
\item We demonstrate the variance-reduction performance of our approach through a detailed simulation study covering two derivative-pricing examples: a dual-condition constant-maturity-swap (CMS) range accrual in a two-factor Gaussian short-rate model (G2++), and a worst-of autocallable with downside protection in a Wishart multivariate stochastic-volatility model.
\end{enumerate}

The remainder of the paper is organised as follows.
\Cref{sec:problem_formulation} introduces the general setting and formulates optimal stratification as a stochastic optimisation problem over neural stratification factors.
\Cref{sec:approximation} establishes continuity of the associated conditional expectations and conditional distributions and derives the universal approximation result for neural stratification factors.
\Cref{sec:simulation_study} examines the variance-reduction performance of neural stratification factors in two benchmark examples.
\Cref{sec:conclusion} summarises the main findings and discusses directions for future research.

\section{Optimal neural stratification}\label{sec:problem_formulation}
We fix $d,n\in\mathbb N$ and a filtered probability space $(\Omega,\mathcal F,\mathbb F,\Prob)$, with $\mathbb F=(\mathcal F_t)_{t\in[0,1]}$, satisfying the usual conditions.
We assume that processes and random variables are, unless stated otherwise, $\mathbb F$-adapted resp.\ $\mathcal F$-measurable and identified up to indistinguishability resp.\ up to $\Prob$-almost-sure equality.
Likewise, sub-$\sigma$-fields are identified whenever their $\Prob$-completions coincide.
Moreover, conditional expectations and random conditional distributions are understood as fixed measurable versions and are identified $\Prob$-almost surely.
Let $M$ be an $\R^d$-valued continuous local martingale with $M_0=0$ and deterministic matrix-valued quadratic covariation process $[M]$, and let $A:[0,1]\to\R$ be deterministic, continuous and of finite variation with $A_0=0$ and corresponding total-variation process $|A|$.
For $r\in\mathbb N$, we write $\mathcal C^r\coloneqq C([0,1];\R^r)$, equipped with the topology of uniform convergence and its Borel $\sigma$-field.
We fix $x_0\in\R^n$, and let $a:[0,1]\times\mathcal C^n\to\R^n$ and $b:[0,1]\times\mathcal C^n\to\R^{n\times d}$ be deterministic, Borel measurable, and non-anticipative functionals, the latter meaning that $a_t(x)\coloneqq a(t,x)$ and $b_t(x)\coloneqq b(t,x)$ satisfy $a_t(x)=a_t(x_{\cdot\wedge t})$ and $b_t(x)=b_t(x_{\cdot\wedge t})$.
We consider continuous adapted solutions to
\begin{equation}\label{eq:sde}
X_t=x_0+\int_0^t a_s(X)\,\mathrm dA_s+\int_0^t b_s(X)\,\mathrm dM_s,
\qquad t\in[0,1],
\end{equation}
for which $a(X)$ is progressively measurable with $\int_0^1\norm{a_s(X)}\,\mathrm d|A|_s<\infty$ almost surely, and $b(X)$ is predictable and stochastically integrable with respect to $M$.
We assume the existence of a strong solution in the sense of \citet[Def.~1.2]{kurtz14weak}, meaning there is a Borel measurable map $\Phi:\mathcal C^d\to\mathcal C^n$ with $X=\Phi(M)$ almost surely, as well as pathwise uniqueness, meaning that any two solutions driven by the same $M$ are indistinguishable.
Finally, let $H:\mathcal C^n\to\R$ be Borel measurable, set $G\coloneqq H(X)$, assume $G\in L^2(\Prob)$, and write $V\coloneqq\E[G]$.

\begin{example}[Generalized geometric model]
Let $n=1$, $\beta\in\R^d$ and $\alpha:[0,1]\to\R$ be Borel measurable with $\int_0^1\abs{\alpha(s)}\,\mathrm d|A|_s<\infty$, and define $a_t(x)\coloneqq\alpha(t)x_t$ and $b_t(x)\coloneqq x_t\beta^\top$, which are Borel measurable and non-anticipative by construction.
We consider the candidate solution
\begin{equation*}
X_t = x_0\,\exp\left( \int_0^t\alpha(s)\,\mathrm dA_s + \beta^\top M_t - \frac12 \beta^\top [M]_t\beta \right).
\end{equation*}
Set $B_t\coloneqq\int_0^t\alpha(s)\,\mathrm dA_s$, $N_t\coloneqq\beta^\top M_t$, $Q_t\coloneqq\beta^\top [M]_t\beta$, and $\mathcal E_t\coloneqq\exp(B_t+N_t-\frac12Q_t)$ so that $X=x_0\mathcal E$.
Because $A$ is continuous, $B$ is continuous and of finite variation, while $N$ is a continuous local martingale with $[N]=Q$.
It\^o's formula therefore yields $\mathrm d\mathcal E_t=\alpha(t)\mathcal E_t\,\mathrm dA_t+\mathcal E_t\beta^\top\,\mathrm dM_t$, hence $x_0\mathcal E$ satisfies \cref{eq:sde}.
It\^o's formula moreover yields $\mathrm d(\mathcal E_t^{-1})=\mathcal E_t^{-1}(-\alpha(t)\,\mathrm dA_t-\beta^\top\,\mathrm dM_t+\mathrm dQ_t)$, while for any other solution $\widetilde X$, $\mathrm d[\widetilde X,\mathcal E^{-1}]_t=-\widetilde X_t\mathcal E_t^{-1}\,\mathrm dQ_t$, and integration by parts therefore implies $\mathrm d(\widetilde X_t\mathcal E_t^{-1})=0$.
Since $\widetilde X_0=x_0$ and $\mathcal E_0=1$, this proves $\widetilde X=x_0\mathcal E$ and hence pathwise uniqueness.
Moreover, since $B$ and $Q$ are fixed continuous functions, the map $\Phi:\mathcal C^d\to\mathcal C^1$ defined by $\Phi(m)_t\coloneqq x_0\exp(B_t+\beta^\top m_t-\frac12Q_t)$ is continuous for the uniform topology and satisfies $X=\Phi(M)$.
Taking $d=1$, $A_t=t$, $M$ to be a standard Brownian motion, $\alpha\equiv\mu$ for $\mu\in\R$, $\beta=\sigma>0$, and $x_0>0$ recovers the geometric Brownian motion as a special case.
\end{example}

To construct a stratified estimator of $V$, let $q,K$ be positive integers and let $Y$ be an $\R^q$-valued stratification factor.
Choose a finite Borel partition $\mathcal D\coloneqq\{D_1,\ldots,D_K\}$ of $\R^q$, define the $\{1,\ldots,K\}$-valued stratum index $S_Y\coloneqq\sum_{k=1}^K k\,\mathbf 1_{\{Y\in D_k\}}$, set $p_k\coloneqq\Prob(S_Y=k)=\Prob(Y\in D_k)$ for $k=1,\ldots,K$, and assume that $p_k > 0$ for every $k$.
For positive integers $n_1,\ldots,n_K$, write $\mathbf n\coloneqq(n_1,\ldots,n_K)$ and let the $\mathcal C^n$-valued random elements $X^{(k,\ell)}$, $k=1,\ldots,K$ and $\ell=1,\ldots,n_k$, be mutually independent with $X^{(k,\ell)}\sim\mathcal L(X\mid S_Y=k)$.
We define the total sample size by $N\coloneqq\sum_{k=1}^K n_k$, and the stratified estimator of $V$ by
\begin{equation*}
\widehat V_{\mathcal D,\mathbf n}\coloneqq\sum_{k=1}^K p_k\widehat V_k,
\qquad
\widehat V_k\coloneqq\frac{1}{n_k}\sum_{\ell=1}^{n_k}H\bigl(X^{(k,\ell)}\bigr).
\end{equation*}
By construction, $\E[\widehat V_k]=\E[G\mid S_Y=k]$ for each $k$, and the law of total expectation yields $\E[\widehat V_{\mathcal D,\mathbf n}] = V$, so the stratified estimator is unbiased.

For $k=1,\ldots,K$, set $\sigma_k\coloneqq\sqrt{\Var(G\mid S_Y=k)}$.
Then $\sigma_k <\infty$ for every $k$, and $\Var(\widehat V_k)=\sigma_k^2/n_k$.
Let $\mathbf n(N)$ be a sequence of allocations such that $\sum_{k=1}^K n_k(N)=N$ and $n_k(N)/N\to\alpha_k>0$ for each $k$.
Then, under proportional allocation with $\alpha_k=p_k$,
\begin{equation}\label{eq:asymptotic_variance}
\lim_{N\to\infty}N\Var\bigl(\widehat V_{\mathcal D,\mathbf n(N)}\bigr)
=\sum_{k=1}^K\frac{p_k^2\sigma_k^2}{\alpha_k} = \E\bigl[\Var(G\mid\sigma(S_Y))\bigr]=\Var(G)-\Var\bigl(\E[G\mid\sigma(S_Y)]\bigr),
\end{equation}
where the third identity follows from the law of total variance.
From this we can see that, asymptotically and under proportional allocation, the variance of the stratified estimator is reduced by $\Var(\E[G\mid\sigma(S_Y)])/N$ relative to crude Monte Carlo, which has variance $\Var(G)/N$.
For general allocation proportions, the first equality in \cref{eq:asymptotic_variance} remains valid.
Assuming all the within-stratum variances $\sigma_k^2$ to be strictly positive, and writing $s\coloneqq\sum_{k=1}^Kp_k\sigma_k$, the Cauchy--Schwarz inequality implies $s^2\leq\sum_{k=1}^Kp_k^2\sigma_k^2/\alpha_k$, with equality exactly for $\alpha_k^\ast=p_k\sigma_k/s$.
The minimal asymptotic scaled variance is therefore given by $s^2$, and the corresponding optimal allocation is $\alpha_k^\ast$.
However, since for $\alpha_k \neq p_k$, the connection to the explained variance $\Var(\E[G\mid\sigma(S_Y)])$ is less direct, we focus on the identification of optimal stratification factors under proportional allocation.

We now adapt \citet[Lemma~4.1]{glasserman99asymptotically} to our setting, which allows us to characterise, asymptotically, the variance of stratified estimators in terms of $\Var(\E[G\mid\sigma(Y)])$ instead of $\Var(\E[G\mid\sigma(S_Y)])$.
Rather than working with a fixed partition $\mathcal D$, let $(\mathcal D_r)_{r\in\mathbb N}$, with $\mathcal D_r=\{D_{r,1},\ldots,D_{r,K_r}\}$, be finite Borel partitions of $\R^q$ such that every cell of $\mathcal D_{r+1}$ is contained in a cell of $\mathcal D_r$.
We set $p_{r,k}\coloneqq\Prob(Y\in D_{r,k})>0$ and define $S_r\coloneqq\sum_{k=1}^{K_r}k\,\mathbf 1_{\{Y\in D_{r,k}\}}$ and $\mathcal G_r\coloneqq\sigma(S_r)$.
For every $r$, let $\mathbf n_r(N)$ be allocations with positive integer components, such that $\sum_{k=1}^{K_r}n_{r,k}(N)=N$ and $n_{r,k}(N)/N\to p_{r,k}$, and write $\widehat V_{r,N}\coloneqq\widehat V_{\mathcal D_r,\mathbf n_r(N)}$.
\begin{lemma}\label{lem:fine_stratification}
Assume that $\mathcal G_\infty\coloneqq\sigma(\bigcup_{r\in\mathbb N}\mathcal G_r) = \sigma(Y)$.
Then
\begin{equation*}
\lim_{r\to\infty}\lim_{N\to\infty}N\Var(\widehat V_{r,N})
=\E\bigl[\Var(G\mid\sigma(Y))\bigr]
=\Var(G)-\Var(m_Y).
\end{equation*}
\end{lemma}
The main assumption of \cref{lem:fine_stratification} is sufficient for identifying the limit from the martingale convergence theorem with $\E[G\mid\sigma(Y)]$.
Compared with the setting of \citet[Lemma~4.1]{glasserman99asymptotically}, the present lemma allows arbitrary $\R^q$-valued factors instead of real-valued Gaussian factors, and arbitrary nested finite Borel partitions instead of equiprobable quantile intervals.
In the setting of \citet{glasserman99asymptotically}, the main assumption of \cref{lem:fine_stratification} is automatically satisfied by construction, whereas in the present setting we need to impose it explicitly.
However, in \cref{sec:simulation_study}, we will work with an adaptation of the stratification design of \citet{glasserman99asymptotically}, which ensures that the assumption is satisfied.
\begin{proof}[Proof of \cref{lem:fine_stratification}]
For fixed $r$, \cref{eq:asymptotic_variance} yields $\lim_{N\to\infty}N\Var(\widehat V_{r,N}) = \E[\Var(G\mid\mathcal G_r)]$.
The partitions are nested, hence $(\mathcal G_r)_{r\in\mathbb N}$ is increasing.
Since $G \in L^2(\Prob)$, the martingale convergence theorem implies $\E[G\mid\mathcal G_r]\to\E[G\mid\mathcal G_\infty]=m_Y$ in $L^2(\Prob)$.
By the law of total variance, $\E[\Var(G\mid\mathcal G_r)]=\Var(G)-\Var(\E[G\mid\mathcal G_r])\to\Var(G)-\Var(m_Y)=\E[\Var(G\mid\sigma(Y))]$.
\end{proof}

The deterministic covariation $[M]$ induces a unique finite, positive-semidefinite and matrix-valued measure $C$ on $[0,1]$ via $C((s,t])=[M]_t-[M]_s$ for $0\leq s<t\leq1$ and $C(\{0\})=0$.
Since $\operatorname{tr}[M]_1<\infty$, an application of the Burkholder--Davis--Gundy inequality yields $\E[\sup_{t\in[0,1]}\norm{M_t}^2]<\infty$, so $M$ is a square-integrable martingale.
We define the finite trace measure $\mu(B)\coloneqq\operatorname{tr}C(B)$ on $\mathcal B([0,1])$, for which $\mu([0,1])=\operatorname{tr}[M]_1$.
Applying the Radon--Nikodym theorem componentwise to $C$ with respect to $\mu$ yields a measurable symmetric matrix-valued function $\pi$ satisfying
\begin{equation*}
C(B)=\int_B\pi(t)\,\mu(\mathrm dt), \qquad B\in\mathcal B([0,1]).
\end{equation*}
We consider the following space of deterministic integrands:
\begin{equation*}
\mathcal H\coloneqq\left\{f:[0,1]\to\R^d\text{ Borel measurable}:\int_0^1 f(t)^\top\pi(t)f(t)\,\mu(\mathrm dt)<\infty\right\}.
\end{equation*}
On $\mathcal H$ we define the equivalence relation $f\sim g$ if and only if $\int_0^1(f(t)-g(t))^\top\pi(t)(f(t)-g(t))\,\mu(\mathrm dt)=0$.
We denote by $\Lambda^2 \coloneqq\mathcal H /\!\sim$ the quotient space which, endowed with the inner product $\langle f,g\rangle_{\Lambda^2}\coloneqq\int_0^1f(t)^\top\pi(t)g(t)\,\mu(\mathrm dt)$, is a separable Hilbert space.
The space $\Lambda^2$ is a special case of Jacod's construction of the vector stochastic integral. In the multidimensional setting, the space of componentwise integrable deterministic integrands need not be complete under the seminorm induced by the covariation of $M$, and $\Lambda^2$ provides its natural completion; see \citet[Chap.~IV]{jacod79calcul} and \citet{shiryaev02vector} for further details.

For $f\in\Lambda^2$, we define the stochastic integral process $\mathcal I_\cdot(f)$ and its terminal value by
\begin{equation*}
\mathcal I_t(f)\coloneqq\int_0^t f(s)^\top\,\mathrm dM_s,
\qquad
\mathcal I(f)\coloneqq\mathcal I_1(f),
\end{equation*}
where these two quantities are independent of the chosen representative of $f$ up to indistinguishability and $\Prob$-almost sure equality, respectively.
In particular, $\mathcal I:\Lambda^2\to L^2(\Prob)$ is a linear isometry.
More generally, for $F=(f_1,\ldots,f_q)\in(\Lambda^2)^q$, we regard $F(t)$ as the $d\times q$ matrix with columns $f_1(t),\ldots,f_q(t)$ and set
\begin{equation*}
Y^F\coloneqq\int_0^1F(t)^\top\,\mathrm dM_t =\bigl(\mathcal I(f_1),\ldots,\mathcal I(f_q)\bigr)^\top,
\qquad
\Sigma_F\coloneqq\Cov(Y^F) =\bigl(\langle f_i,f_j\rangle_{\Lambda^2}\bigr)_{i,j=1}^q.
\end{equation*}
\citet[Remarks~2.4 and~4.3]{arandjelovic25importance} show that $\mathcal I(f)$ is centred Gaussian for every $f\in\Lambda^2$ and, since $a^\top Y^F=\mathcal I(Fa)$ for every $a\in\R^q$, it follows that $Y^F\sim\mathcal N(0,\Sigma_F)$.
Moreover, $a^\top\Sigma_Fa=\norm{Fa}_{\Lambda^2}^2$, so $Y^F$ is non-degenerate if and only if $\Sigma_F$ is positive definite, or equivalently if and only if $f_1,\ldots,f_q$ are linearly independent in $\Lambda^2$.
For such $F$, letting $\Sigma_F^{-1/2}$ denote the symmetric positive definite inverse square root and setting $\overline F\coloneqq F\Sigma_F^{-1/2}$ gives $Y^{\overline F}=\Sigma_F^{-1/2}Y^F\sim\mathcal N(0,I_q)$, and invertibility of $\Sigma_F^{-1/2}$ implies that $\sigma(Y^{\overline F})$ and $\sigma(Y^F)$ coincide.
Finally, if $F_n=(f_{n,1},\ldots,f_{n,q})\to F$ in $(\Lambda^2)^q$, then the It\^o isometry yields $\E[\norm{Y^{F_n}-Y^F}^2]=\sum_{j=1}^q\norm{f_{n,j}-f_j}_{\Lambda^2}^2\to0$.

We now have all the ingredients to formulate optimal stratification as a stochastic optimisation problem over neural stratification factors.
Let $F_\theta=(f_{\theta,1},\ldots,f_{\theta,q}):[0,1]\to\R^{d\times q}$, $\theta\in\Theta$, have columns given by $\R^d$-valued continuous feedforward neural networks.
Each $F_\theta$ is bounded on $[0,1]$, and its columns therefore represent elements of $\Lambda^2$, since, for $j=1,\ldots,q$,
\begin{equation*}
\int_0^1 f_{\theta,j}(t)^\top\pi(t)f_{\theta,j}(t)\,\mu(\mathrm dt)\leq \sup_{t\in[0,1]}\norm{f_{\theta,j}(t)}^2\operatorname{tr}[M]_1<\infty.
\end{equation*}
We write $\Theta_\mathrm{nd}\coloneqq\{\theta\in\Theta:\Sigma_{F_\theta}\text{ is positive definite}\}$ and assume $\Theta_{\mathrm{nd}}\neq\varnothing$.
For $\theta\in\Theta_{\mathrm{nd}}$, set $\overline F_\theta\coloneqq F_\theta\Sigma_{F_\theta}^{-1/2}$ and $Y_\theta\coloneqq Y^{\overline F_\theta}$.
In view of \cref{lem:fine_stratification}, optimal neural stratification factors are defined as solutions to the problem
\begin{equation}\label{eq:optimization_problem}
\sup_{\theta\in\Theta_{\mathrm{nd}}} \Var(\E[ G\mid \sigma(Y_\theta)]) = \Var(G) - \inf_{\theta\in\Theta_{\mathrm{nd}}} \E[\Var(G\mid\sigma(Y_\theta))].
\end{equation}

\begin{remark}
The objective in \cref{eq:optimization_problem} is in general neither convex nor concave, and may admit suboptimal local maxima.
These features are already present in the simplest Gaussian setting.
Let $q=1$, let $M=(W^1,W^2)$ be a two-dimensional standard Brownian motion, and let $e_1,e_2\in\Lambda^2$ denote the corresponding constant coordinate integrands.
Put $Z_i\coloneqq\mathcal I(e_i)=W_1^i$, $i=1,2$, so that $Z_1$ and $Z_2$ are independent standard normal random variables.

First take $G=Z_1$ and, for $b\in\mathbb R$, let $f_b=e_1+be_2$.
Then $Y^{f_b}=Z_1+bZ_2$, and Gaussian regression yields
\begin{equation*}
\E[G\mid\sigma(Y^{f_b})]=\frac{1}{1+b^2}Y^{f_b},\qquad J(b)\coloneqq\Var\!\left(\E[G\mid\sigma(Y^{f_b})]\right)=\frac{1}{1+b^2}.
\end{equation*}
Since $J(0)=1>(J(-1)+J(1))/2=1/2$, the objective is not convex.
On the other hand, $J(1)=1/2<(J(0)+J(2))/2=3/5$, so it is not concave.

Strict local maxima have to be understood after removing the immaterial scaling of the factor, since $\sigma(Y^{cf})=\sigma(Y^f)$ for every $c\neq0$.
To see that suboptimal strict local maxima can nevertheless occur on the space of one-dimensional factor directions, now let $G=2(Z_1^2-1)-(Z_2^2-1)$.
For an arbitrary nonzero $f\in\Lambda^2$, write $f=ae_1+be_2+h$, where $h$ is orthogonal to $e_1$ and $e_2$, and set $r^2\coloneqq\norm{h}_{\Lambda^2}^2$ and $s^2\coloneqq\norm{f}_{\Lambda^2}^2=a^2+b^2+r^2$.
Then $\mathcal I(h)$ is independent of $(Z_1,Z_2)$ and $Y^f=aZ_1+bZ_2+\mathcal I(h)$ is centred Gaussian with variance $s^2$.
Gaussian regression therefore gives
\begin{equation*}
\E[Z_1^2-1\mid\sigma(Y^f)]=\frac{a^2}{s^2}\left(\frac{(Y^f)^2}{s^2}-1\right),\qquad \E[Z_2^2-1\mid\sigma(Y^f)]=\frac{b^2}{s^2}\left(\frac{(Y^f)^2}{s^2}-1\right).
\end{equation*}
Since $(Y^f)/s$ is standard normal, it follows that
\begin{equation}\label{eq:suboptimal_local_maximum}
\Var\!\left(\E[G\mid\sigma(Y^f)]\right)=2\left(\frac{2a^2-b^2}{a^2+b^2+r^2}\right)^2.
\end{equation}
By scale invariance, restrict to $\norm{f}_{\Lambda^2}=1$.
Then \cref{eq:suboptimal_local_maximum} becomes $2(-1+3a^2+r^2)^2$.
At the direction generated by $e_2$, this value equals $2$.
For unit factors whose directions converge to the direction generated by $e_2$, one has $a^2+r^2=1-b^2\to0$.
If the direction is distinct from that generated by $e_2$, then $a^2+r^2>0$, and hence $0<3a^2+r^2<2$ for all sufficiently close directions.
Therefore $2(-1+3a^2+r^2)^2<2$.
Thus the direction generated by $e_2$ is a strict local maximiser among all one-dimensional factor directions in $\Lambda^2$.
It is not a global maximiser, since the direction generated by $e_1$ yields the value $8$.

These observations preclude a direct convex-optimisation treatment of \cref{eq:optimization_problem} and motivate the numerical approach adopted below: neural parametrisations combined with stochastic-gradient methods provide a flexible means of searching over a rich class of stratification factors without imposing convexity or concavity on the objective.
\end{remark}

\begin{remark}
The restriction to $[0,1]$ is made for notational convenience.
Let $T>0$ and suppose that the model is originally formulated on $[0,T]$, with the corresponding objects denoted by tildes.
For a path $x$ on $[0,1]$, write $x^T_s:=x_{s/T}$, $s\in[0,T]$, and define, for $t\in[0,1]$,
\begin{equation*}
X_t\coloneqq\widetilde X_{Tt},\qquad
M_t\coloneqq\widetilde M_{Tt},\qquad
A_t\coloneqq\widetilde A_{Tt},\qquad
a_t(x)\coloneqq\widetilde a_{Tt}(x^T),\qquad
b_t(x)\coloneqq\widetilde b_{Tt}(x^T),
\end{equation*}
as well as $H(x)\coloneqq\widetilde H(x^T)$.
Then $[M]_t=[\widetilde M]_{Tt}$, the transformed coefficients remain Borel measurable and non-anticipative, and the time-changed process $X$ satisfies \cref{eq:sde}.
Moreover, $H(X)=\widetilde H(\widetilde X)$, and the corresponding solution map is obtained by the same rescaling.
If $\widetilde C$ and $\widetilde\Lambda^2$ denote the covariation measure and Hilbert space on $[0,T]$, then, for $\widetilde f\in\widetilde\Lambda^2$, setting $f(t)\coloneqq\widetilde f(Tt)$ yields
\begin{equation*}
\norm{f}_{\Lambda^2}=\norm{\widetilde f}_{\widetilde\Lambda^2},
\qquad
\int_0^1 f(t)^\top\,\mathrm dM_t = \int_0^T \widetilde f(s)^\top\,\mathrm d\widetilde M_s,
\end{equation*}
where the identity of norms follows from $C(B)=\widetilde C(TB)$ for $B\in\mathcal B([0,1])$, with $TB:=\{Tt:t\in B\}$.
The same transformation applies columnwise to $F_\theta$, and therefore preserves the resulting stratification factor, its covariance, and the optimisation problem in \cref{eq:optimization_problem}.
\end{remark}

Before we conclude this section, we discuss how to sample from the conditional distribution of $H(X)$ given $Y^F$.
Let $F\in(\Lambda^2)^q$ with $\Sigma_F=I_q$, and define the cross-covariance function $K_F(t)\coloneqq\Cov(M_t,Y^F)=\int_0^t \pi(s)F(s)\,\mu(\mathrm ds)$ for $t\in[0,1]$.
Let $\widetilde M$ be an independent copy of $M$, set $\widetilde Y^F\coloneqq\int_0^1F(t)^\top\,\mathrm d\widetilde M_t$.
The process $R_t^F\coloneqq \widetilde M_t-K_F(t)\widetilde Y^F$ is then independent of $\widetilde Y^F$.
Define
\begin{equation*}
M_t^{F,y}\coloneqq R_t^F+K_F(t)y=\widetilde M_t+K_F(t)(y-\widetilde Y^F),\qquad t\in[0,1], \quad y\in\R^q.
\end{equation*}
Writing $\nu_F\coloneqq\mathcal L(R^F)$, for $A\in\mathcal B(\mathcal C^d)$ set
\begin{equation*}
Q_F(y,A)\coloneqq\Prob(M^{F,y}\in A)=\int_{\mathcal C^d}\mathbf 1_A\!\left(r+K_F(\cdot)y\right)\,\nu_F(\mathrm dr).
\end{equation*}
The map $(r,y)\mapsto r+K_F(\cdot)y$ from $\mathcal C^d\times\R^q$ to $\mathcal C^d$ is continuous, so the integrand is jointly Borel measurable and $y\mapsto Q_F(y,A)$ is Borel measurable; hence $Q_F$ is a probability kernel from $\R^q$ to $\mathcal C^d$.
Then, for $A\in\mathcal B(\mathcal C^d)$ and $B\in\mathcal B(\R^q)$, independence of $R^F$ and $\widetilde Y^F$, together with $(\widetilde M,\widetilde Y^F) \overset{\mathrm{d}}{=} (M,Y^F)$ and $\widetilde M=R^F+K_F(\cdot)\widetilde Y^F$, yields
\begin{equation*}
\Prob(M\in A,Y^F\in B)=\int_B\int_{\mathcal C^d}\mathbf 1_A\!\left(r+K_F(\cdot)y\right)\,\nu_F(\mathrm dr)\,\gamma_q(\mathrm dy)=\int_B Q_F(y,A)\,\gamma_q(\mathrm dy),
\end{equation*}
where $\gamma_q=\mathcal N(0,I_q)$.
Hence, by \citet[Theorem~8.5]{kallenberg21foundations}, $Q_F(Y^F,\cdot)$ is a version of $\mathcal L(M\mid Y^F)$.
Since $X=\Phi(M)$, define $Q_F^H(y,\cdot)\coloneqq\mathcal L(H(\Phi(M^{F,y})))$.
Then $Q_F^H(Y^F,\cdot)$ is a version of $\mathcal L(H(X)\mid Y^F)$.
Hence, for a stratum $D_k$ with $\gamma_q(D_k)>0$,
\begin{equation*}
\mathcal L(H(X)\mid Y^F\in D_k)=\int_{D_k}Q_F^H(y,\cdot)\,\gamma_q(\mathrm dy\mid D_k),
\end{equation*}
so a draw from this distribution is obtained by first drawing $y\sim\gamma_q(\,\cdot\mid D_k)$ and, conditionally on $y$, independently generating $\widetilde M$, constructing $M^{F,y}$, and evaluating $H(\Phi(M^{F,y}))$.

\section{Approximation of conditional distributions}\label{sec:approximation}
In order to justify replacing a stratification factor by an approximation from the class of neural factors in \cref{eq:optimization_problem}, we need to understand how perturbations of its deterministic integrands affect the associated conditional quantities.
For $q\in\mathbb N$ and $F\in(\Lambda^2)^q$, we write $\mathcal A_F\coloneqq\sigma(Y^F)$.
If $F_n\to F$ in $(\Lambda^2)^q$, then $Y^{F_n}\to Y^F$ in $L^2(\Prob)$, and the central question is whether, and in which sense, $m_{Y^{F_n}}$ and $\mu_{Y^{F_n}}$ converge to $m_{Y^F}$ and $\mu_{Y^F}$, respectively.
Note that the $\sigma$-fields $\mathcal A_{F_n}$ need neither be increasing nor decreasing.
The following example shows why the martingale convergence theorem does not by itself resolve the perturbation problem.

\begin{example}\label{ex:martingale_convergence}
Let $d=n=2$ and let $M=(B^1,B^2)$ be a two-dimensional standard Brownian motion.
Take $x_0=0$, $a\equiv0$ and $b\equiv I_2$, so that $X=M$, and set $H(x)\coloneqq x_1^{(1)}$.
With $f(t)\equiv e_1$ and $g(t)\equiv e_2$, which are orthonormal in $\Lambda^2$, write $Z\coloneqq\mathcal I(f)=B_1^1$ and $W\coloneqq\mathcal I(g)=B_1^2$, which are independent by construction, and set $G=Z$ and $\varepsilon_n\coloneqq n^{-1}$.
For $\delta\in\{0,1\}$, define $F_n^{(\delta)},F^{(\delta)}\in(\Lambda^2)^2$ and the corresponding factors $Y_n^{(\delta)}\coloneqq Y^{F_n^{(\delta)}}$ and $Y^{(\delta)}\coloneqq Y^{F^{(\delta)}}$ by $F^{(\delta)}\coloneqq(\delta f,\delta g)$, $F_n^{(\delta)}\coloneqq\bigl(\delta f,\varepsilon_n f+(\delta+\varepsilon_n^2)g\bigr)$, and
\begin{equation*}
Y_n^{(\delta)}=
\begin{pmatrix}
\delta Z\\
\varepsilon_n Z+(\delta+\varepsilon_n^2)W
\end{pmatrix},
\qquad
Y^{(\delta)}=\delta
\begin{pmatrix}
Z\\ W
\end{pmatrix},
\end{equation*}
and note that $\norm{F_n^{(\delta)}-F^{(\delta)}}_{(\Lambda^2)^2}^2=\E\![\norm{Y_n^{(\delta)}-Y^{(\delta)}}^2]=\varepsilon_n^2+\varepsilon_n^4\rightarrow0$.
For $\delta=1$, the matrix $\left(\begin{smallmatrix}1&0\\ \varepsilon_n&1+\varepsilon_n^2\end{smallmatrix}\right)$ has determinant $1+\varepsilon_n^2>0$, and hence $\sigma(Y_n^{(1)})=\sigma(Z,W)=\sigma(Y^{(1)})$ for every $n$.
The conditioning $\sigma$-fields are therefore constant in $n$, so the $L^2$ martingale convergence theorem applies.
In fact, $\E[G\mid\sigma(Y_n^{(1)})]=Z=\E[G\mid\sigma(Y^{(1)})]$ for every $n$.

For $\delta=0$, since $\varepsilon_n>0$, one has $\sigma(Y_n^{(0)})=\sigma(Z+\varepsilon_nW)$, whereas $\sigma(Y^{(0)})$ is $\Prob$-trivial.
If $m\neq n$, one has
\begin{equation*}
\Var\!\left(Z+\varepsilon_nW\mid\sigma(Z+\varepsilon_mW)\right)=1+\varepsilon_n^2-\frac{(1+\varepsilon_n\varepsilon_m)^2}{1+\varepsilon_m^2}=\frac{(\varepsilon_n-\varepsilon_m)^2}{1+\varepsilon_m^2}>0.
\end{equation*}
Thus $Z+\varepsilon_nW$ is not measurable with respect to $\sigma(Z+\varepsilon_mW)$ (for otherwise the conditional variance would be zero), and interchanging $m$ and $n$ shows that the conditioning $\sigma$-fields are pairwise non-nested.
Consequently, neither the martingale nor the reverse-martingale convergence theorem applies.
In fact, $\Cov(Z,Z+\varepsilon_nW)=1$, $\Var(Z+\varepsilon_nW)=1+\varepsilon_n^2$, and
\begin{equation*}
\E[G\mid\sigma(Y_n^{(0)})]=\frac{Z+\varepsilon_nW}{1+\varepsilon_n^2},
\qquad
\E[G\mid\sigma(Y^{(0)})]=\E[G]=0,
\end{equation*}
hence $\norm{\E[G\mid\sigma(Y_n^{(0)})]-\E[G\mid\sigma(Y^{(0)})]}_{L^2(\Prob)}=(1+\varepsilon_n^2)^{-1/2}\rightarrow1$, so the sequence of conditional expectations $\E[G\mid\sigma(Y_n^{(0)})]$ does not converge to $\E[G\mid\sigma(Y^{(0)})]$.
This example shows that neither convergence of the integrands nor $L^2$-convergence of the Gaussian factors need produce a monotone sequence of conditioning $\sigma$-fields.
\end{example}

The failure of the generated $\sigma$-fields to be nested in \cref{ex:martingale_convergence} is explained by a simple property of one-dimensional Gaussian factors, which we record in \cref{lem:gaussian_factor_collinearity}.
For $0\neq f\in\Lambda^2$, set $\widehat f\coloneqq f/\norm{f}_{\Lambda^2}$ and $\mathcal A_f\coloneqq\sigma(\mathcal I(f))$.
We call a sequence $(f_n)_{n\in\mathbb N}$ in $\Lambda^2\setminus\{0\}$ eventually collinear if there exist $m\in\mathbb N$ and $u\in\Lambda^2$ with $\norm{u}_{\Lambda^2}=1$ such that $\widehat f_n\in\{u,-u\}$ for every $n\geq m$.

\begin{lemma}\label{lem:gaussian_factor_collinearity}
Let $f,g\in\Lambda^2\setminus\{0\}$.
If $\widehat f=\pm\widehat g$, then $\mathcal A_f=\mathcal A_g$; otherwise $\mathcal A_f\cap\mathcal A_g$ is $\Prob$-trivial.
Moreover, if $(f_n)_{n\in\mathbb N}$ is not eventually collinear, then $\bigvee_{m\geq1}\bigcap_{n\geq m}\mathcal A_{f_n}$ is $\Prob$-trivial.
\end{lemma}

\begin{proof}
Set $X\coloneqq\mathcal I(f)$, $Y\coloneqq\mathcal I(g)$, $\sigma_X\coloneqq\norm{f}_{\Lambda^2}$, $\sigma_Y\coloneqq\norm{g}_{\Lambda^2}$ and $\rho\coloneqq\langle\widehat f,\widehat g\rangle_{\Lambda^2}$.
Then $(X,Y)$ is centred Gaussian, with variances $\sigma_X^2$ and $\sigma_Y^2$ and covariance $\rho\sigma_X\sigma_Y$.
If $\widehat f=\pm\widehat g$, linearity of $\mathcal I$ gives $X=\pm(\sigma_X/\sigma_Y)Y$, and hence $\mathcal A_f=\mathcal A_g$.

Suppose now that $\widehat f\neq\pm\widehat g$.
Then $\abs{\rho}<1$ by the Cauchy--Schwarz inequality.
Let $E\in\mathcal A_f\cap\mathcal A_g$.
Then there exist $B_f,B_g\in\mathcal B(\R)$ such that $\mathbf 1_E=\mathbf 1_{B_f}(X)=\mathbf 1_{B_g}(Y)$ almost surely.
Define
\begin{equation*}
a\coloneqq\rho\frac{\sigma_X}{\sigma_Y},\qquad
b\coloneqq\sigma_X\sqrt{1-\rho^2},\qquad
U\coloneqq\frac{X-aY}{b}.
\end{equation*}
Here $b>0$, $(U,Y)$ is jointly Gaussian, and
\begin{equation*}
\Cov(U,Y)=\frac{\rho\sigma_X\sigma_Y-a\sigma_Y^2}{b}=0,
\qquad
\Var(U)=\frac{\sigma_X^2-2a\rho\sigma_X\sigma_Y+a^2\sigma_Y^2}{b^2}=1.
\end{equation*}
Thus $U$ is standard normal and independent of $Y$, while $X=aY+bU$.
It follows that $\mathbf 1_{B_f}(aY+bU)=\mathbf 1_{B_g}(Y)$ almost surely.
The joint distribution of $(Y,U)$ has a strictly positive density on $\R^2$ and is therefore equivalent to the Lebesgue measure on $\R^2$.
Hence $\mathbf 1_{B_f}(ay+bu)=\mathbf 1_{B_g}(y)$ for Lebesgue-almost every $(y,u)\in\R^2$.
We now define $A\coloneqq \{(y,u)\in\R^2:\mathbf 1_{B_f}(ay+bu)\neq\mathbf 1_{B_g}(y)\}$.
Then $A$ is a Lebesgue-null set.
For each $y\in\R$, let $A_y\coloneqq\{u\in\R:(y,u)\in A\}$.
Tonelli's theorem gives
\begin{equation*}
0 = \int_\R \int_\R \mathbf 1_A(y,u)\,\mathrm du\,\mathrm dy = \int_\R \lambda(A_y)\,\mathrm dy,
\end{equation*}
where $\lambda$ denotes the Lebesgue measure on $\R$.
Since $y \mapsto \lambda(A_y)$ is nonnegative, it must hold that $\lambda(A_y)=0$ for $\lambda$-almost every $y$.
For any such $y$, set $c\coloneqq\mathbf 1_{B_g}(y)$.
Since $b > 0$, the map $u\mapsto ay+bu$ is an affine bijection of $\R$ and maps Lebesgue-null sets to Lebesgue-null sets.
Hence $\mathbf 1_{B_f}(x) = c$ for Lebesgue-almost every $x\in\R$, which implies that either $B_f$ or $\R\setminus B_f$ is a Lebesgue-null set.
Since $X$ is non-degenerate Gaussian, its distribution is absolutely continuous with respect to Lebesgue measure, and hence $\Prob(E)=\Prob(X\in B_f)\in\{0,1\}$.
Since $E\in\mathcal A_f\cap\mathcal A_g$ was arbitrary, $\mathcal A_f\cap\mathcal A_g$ is $\Prob$-trivial.

Finally, suppose that $(f_n)_{n\in\mathbb N}$ is not eventually collinear.
For every $m\in\mathbb N$, there are $n,k\geq m$ such that $\widehat f_n\neq\pm\widehat f_k$.
$\bigcap_{j\geq m}\mathcal A_{f_j}$ is then contained in $\mathcal A_{f_n}\cap\mathcal A_{f_k}$, which is $\Prob$-trivial.
The join $\bigvee_{m\geq1}\bigcap_{n\geq m}\mathcal A_{f_n}$ must therefore be $\Prob$-trivial as well.
\end{proof}

For a sequence $(\mathcal A_n)_{n\in\mathbb N}$, define the lower and upper limits by
\begin{equation*}
\underline{\mathcal A}\coloneqq\bigvee_{m\geq1}\bigcap_{n\geq m}\mathcal A_n,
\qquad
\overline{\mathcal A}\coloneqq\bigcap_{m\geq1}\bigvee_{n\geq m}\mathcal A_n.
\end{equation*}
If $\underline{\mathcal A}=\overline{\mathcal A}$, then, specialising \citet[Theorem~4]{fetter77continuity} to $p=2$ yields $\E[U\mid\mathcal A_n]\rightarrow\E[U\mid\overline{\mathcal A}]$ in $L^2(\Prob)$ for every $U\in L^2(\Prob)$.

\begin{example}\label{ex:fetter_criterion}
We retain the setting of \cref{ex:martingale_convergence}, in particular the notation $G=Z=\mathcal I(f)$, $W=\mathcal I(g)$ and $\varepsilon_n=n^{-1}$, and set $Y\coloneqq Z$.
For the collinear perturbation $Y_n^{\parallel}\coloneqq\mathcal I((1+\varepsilon_n)f)=(1+\varepsilon_n)Z$, one has $\norm{(1+\varepsilon_n)f-f}_{\Lambda^2}=\norm{Y_n^{\parallel}-Y}_{L^2(\Prob)}=\varepsilon_n$ and $\mathcal A_n^{\parallel}\coloneqq\sigma(Y_n^{\parallel})=\sigma(Z)$ for every $n$.
Consequently, $\underline{\mathcal A}^{\parallel}=\overline{\mathcal A}^{\parallel}=\sigma(Y)$, so \citet[Theorem~4]{fetter77continuity} applies.
In fact, $\E[G\mid\mathcal A_n^{\parallel}]=Z=\E[G\mid\sigma(Y)]$ for every $n$.

On the other hand, for the perturbation $Y_n^{\perp}\coloneqq\mathcal I(f+\varepsilon_n g)=Z+\varepsilon_nW$ of $Z$, one likewise has $\norm{f+\varepsilon_n g-f}_{\Lambda^2}=\norm{Y_n^{\perp}-Y}_{L^2(\Prob)}=\varepsilon_n$.
For distinct $n$ and $k$, the normalised directions $u_n\coloneqq(f+\varepsilon_n g)/\sqrt{1+\varepsilon_n^2}$ and $u_k\coloneqq(f+\varepsilon_k g)/\sqrt{1+\varepsilon_k^2}$ are not collinear.
In order to see this, suppose that $u_n = -u_k$.
In that case, $(1+\varepsilon_n^2)^{-1/2} = \langle u_n, f \rangle_{\Lambda^2} = \langle -u_k, f \rangle_{\Lambda^2} = -(1+\varepsilon_k^2)^{-1/2}$, which yields a contradiction.
Now suppose that $u_n = u_k$.
In that case, $\varepsilon_n = \langle u_n, g \rangle_{\Lambda^2} / \langle u_n, f \rangle_{\Lambda^2} = \langle u_k, g \rangle_{\Lambda^2} / \langle u_k, f \rangle_{\Lambda^2} = \varepsilon_k$, whence $n=k$, a contradiction.
The sequence $(f+\varepsilon_n g)_{n\in\mathbb N}$ is therefore not eventually collinear, and \cref{lem:gaussian_factor_collinearity} shows that the lower limit $\underline{\mathcal A}^{\perp}$ of $\mathcal A_n^{\perp}\coloneqq\sigma(Y_n^{\perp})$ is $\Prob$-trivial.
For every $m$ and distinct $n,k\geq m$, one recovers $W=(Y_n^{\perp}-Y_k^{\perp})/(\varepsilon_n-\varepsilon_k)$ and $Z=(\varepsilon_nY_k^{\perp}-\varepsilon_kY_n^{\perp})/(\varepsilon_n-\varepsilon_k)$.
Since every $Y_j^{\perp}$ is $\sigma(Z,W)$-measurable, it follows that $\bigvee_{j\geq m}\mathcal A_j^{\perp}=\sigma(Z,W)$ for every $m$, and hence $\overline{\mathcal A}^{\perp}=\sigma(Z,W)$.
As $\Prob(Z>0)=1/2$, $\overline{\mathcal A}^{\perp}$ is not $\Prob$-trivial, so one has $\underline{\mathcal A}^{\perp}\neq\overline{\mathcal A}^{\perp}$.
Nevertheless, $\Cov(G,Y_n^{\perp})=1$ and $\Var(Y_n^{\perp})=1+\varepsilon_n^2$, and direct computation gives:
\begin{equation*}
\E[G\mid\mathcal A_n^{\perp}]=\frac{Y_n^{\perp}}{1+\varepsilon_n^2},
\qquad
\norm{\E[G\mid\mathcal A_n^{\perp}]-\E[G\mid\sigma(Y)]}_{L^2(\Prob)}=\frac{\varepsilon_n}{\sqrt{1+\varepsilon_n^2}}\rightarrow0.
\end{equation*}
We have thus constructed an explicit example where $\underline{\mathcal A}^{\perp}\neq\overline{\mathcal A}^{\perp}$, but $\E[G\mid\mathcal A_n^{\perp}]\to\E[G\mid\sigma(Y)]$ in $L^2(\Prob)$.
Fetter's condition is therefore sufficient but not necessary for the convergence of conditional expectations, see \cref{thm:conditional_expectation_stability}.
\end{example}

\begin{theorem}\label{thm:conditional_expectation_stability}
Let $q\in\mathbb N$ and let $F_n,F\in(\Lambda^2)^q$ satisfy $F_n\to F$ in $(\Lambda^2)^q$ and $\operatorname{rank}\Sigma_{F_n}=\operatorname{rank}\Sigma_F$ for all sufficiently large $n$.
Then, for every $Z\in L^2(\Prob)$,
\begin{equation*}
\E[Z\mid\mathcal A_{F_n}]\rightarrow\E[Z\mid\mathcal A_F]
\qquad\text{in }L^2(\Prob).
\end{equation*}
\end{theorem}

\begin{proof}
We write $F_n=(f_{n,1},\ldots,f_{n,q})$ and $F=(f_1,\ldots,f_q)$, and set $p\coloneqq\operatorname{rank}\Sigma_F$.
We now distinguish two cases.
First, assume that $p=0$.
Then every column of $F$ is zero in $\Lambda^2$ and the same holds for every column of $F_n$ for all sufficiently large $n$.
Both conditioning $\sigma$-fields $\mathcal A_{F_n}$ and $\mathcal A_F$ are then $\Prob$-trivial.

Next, let us assume that $p\geq1$.
We now show that, without loss of generality, we may assume that $p=q$ in this case.
Choose an index set $I\subseteq\{1,\ldots,q\}$ of cardinality $p$ such that the columns of the subfamily $F^I$ are linearly independent.
For $i,j\in I$, Cauchy--Schwarz gives
\begin{equation*}
\abs{\langle f_{n,i},f_{n,j}\rangle_{\Lambda^2}-\langle f_i,f_j\rangle_{\Lambda^2}}\leq\norm{f_{n,i}-f_i}_{\Lambda^2}\norm{f_{n,j}}_{\Lambda^2}+\norm{f_i}_{\Lambda^2}\norm{f_{n,j}-f_j}_{\Lambda^2}\to0.
\end{equation*}
We now enumerate $I=\{i_1,\ldots,i_p\}$, and denote by $G_n^I$ and $G^I$ the Gram matrices of $F_n^I$ and $F^I$, respectively, i.e., $G_n^I\coloneqq(\langle f_{n,i_k},f_{n,i_l}\rangle_{\Lambda^2})_{k,l=1}^p$ and $G^I\coloneqq(\langle f_{i_k},f_{i_l}\rangle_{\Lambda^2})_{k,l=1}^p$.
The preceding convergence holds entrywise, and hence $\det G_n^I\to\det G^I$, since the determinant is a polynomial in the matrix entries.
For $a=(a_1,\ldots,a_p)^\top\in\R^p$,
\begin{equation*}
a^\top G^I a
=\sum_{k,l=1}^p a_k a_l \langle f_{i_k},f_{i_l}\rangle_{\Lambda^2}
=\left\langle\sum_{k=1}^p a_k f_{i_k},\sum_{l=1}^p a_l f_{i_l}\right\rangle_{\Lambda^2}
= \left\|\sum_{k=1}^p a_k f_{i_k}\right\|_{\Lambda^2}^2,
\end{equation*}
which is strictly positive for $a\neq0$ by linear independence of the columns of $F^I$.
Thus $G^I$ is positive definite and $\det G^I>0$.
It follows that $\det G_n^I>0$ for all sufficiently large $n$, so $G_n^I$ is invertible, and the columns of $F_n^I$ are linearly independent for all sufficiently large $n$.
Moreover, by assumption, $\dim\operatorname{span}\{f_{n,1},\ldots,f_{n,q}\}=p$.
Consequently, the columns of $F_n^I$ form a basis of $\operatorname{span}\{f_{n,1},\ldots,f_{n,q}\}$, so for every $j\in\{1,\ldots,q\}$ there exist deterministic coefficients $c_{n,j,1},\ldots,c_{n,j,p}\in\R$ such that
\begin{equation*}
f_{n,j}=\sum_{k=1}^p c_{n,j,k}f_{n,i_k},
\qquad
\mathcal I(f_{n,j})
=\sum_{k=1}^p c_{n,j,k}\mathcal I(f_{n,i_k}),
\end{equation*}
where the second identity follows from linearity of $\mathcal I$.
Thus every coordinate of $Y^{F_n}$ is $\mathcal A_{F_n^I}$-measurable, which yields $\mathcal A_{F_n}\subseteq\mathcal A_{F_n^I}$, and the reverse inclusion holds because the coordinates of $Y^{F_n^I}$ are among those of $Y^{F_n}$.
We have thus shown that $\mathcal A_{F_n}=\mathcal A_{F_n^I}$ for all sufficiently large $n$, and the same argument applies to $F$: since the columns of $F^I$ are linearly independent and $\operatorname{rank}\Sigma_F=p$, they form a basis of $\operatorname{span}\{f_1,\ldots,f_q\}$, and therefore $\mathcal A_F=\mathcal A_{F^I}$.
After discarding finitely many terms and relabelling the retained columns, it is therefore enough to treat the case $q=p$, in which $\Sigma_F$ and every $\Sigma_{F_n}$ are positive definite.

We now pass to orthonormal coordinates without changing the generated conditioning $\sigma$-fields.
Put $A_n\coloneqq\Sigma_{F_n}^{-1/2}$, $A\coloneqq\Sigma_F^{-1/2}$, $\overline F_n\coloneqq F_nA_n$, and $\overline F\coloneqq FA$.
The preceding entrywise convergence of the Gram matrices implies $\Sigma_{F_n}\to\Sigma_F$ in operator norm, since the matrices have fixed finite dimension.
As $\Sigma_{F_n}$ and $\Sigma_F$ are positive definite, the inverse square roots are well defined; moreover, continuity of the map $B\mapsto B^{-1/2}$ on the cone of positive-definite symmetric matrices gives
\begin{equation*}
A_n=\Sigma_{F_n}^{-1/2}\rightarrow\Sigma_F^{-1/2}=A
\qquad\text{in operator norm}.
\end{equation*}
In particular, $(A_n)$ is bounded in operator norm.
Consequently,
\begin{equation*}
\begin{aligned}
\norm{\overline F_n-\overline F}_{(\Lambda^2)^q}
&=\norm{F_nA_n-FA}_{(\Lambda^2)^q}\\
&\leq\norm{(F_n-F)A_n}_{(\Lambda^2)^q}+\norm{F(A_n-A)}_{(\Lambda^2)^q}\\
&\leq\norm{F_n-F}_{(\Lambda^2)^q}\norm{A_n}_{\mathrm{op}}+\norm{F}_{(\Lambda^2)^q}\norm{A_n-A}_{\mathrm{op}}
\rightarrow0.
\end{aligned}
\end{equation*}
Since $A_n$ and $A$ are symmetric, the Gram matrices of $\overline F_n$ and $\overline F$ are
\begin{equation*}
A_n^\top\Sigma_{F_n}A_n=\Sigma_{F_n}^{-1/2}\Sigma_{F_n}\Sigma_{F_n}^{-1/2}=I_q,
\qquad
A^\top\Sigma_FA=I_q.
\end{equation*}
Thus, writing $\overline F_n=(u_{n,1},\ldots,u_{n,q})$ and $\overline F=(u_1,\ldots,u_q)$, the respective columns form orthonormal families in $\Lambda^2$.
Finally, by linearity of the stochastic integral and symmetry of $A_n$ and $A$,
\begin{equation*}
Y^{\overline F_n}=Y^{F_nA_n}=A_nY^{F_n},
\qquad
Y^{\overline F}=Y^{FA}
=
AY^F.
\end{equation*}
Since $A_n$ and $A$ are deterministic and invertible, each factor is a measurable function of the other, and hence
$\mathcal A_{\overline F_n}=\mathcal A_{F_n}$ and $\mathcal A_{\overline F}=\mathcal A_F$.

Fix $h\in\Lambda^2$ and set $a_{n,j}\coloneqq\langle h,u_{n,j}\rangle_{\Lambda^2}$, $r_n\coloneqq h-\sum_{j=1}^q a_{n,j}u_{n,j}$, with $a_j$ and $r$ defined analogously from $u_j$.
Since $u_{n,j}\to u_j$ for every $j$, one has $a_{n,j}\to a_j$ and $r_n\to r$ in $\Lambda^2$.
The vector $(\mathcal I(r_n),Y^{\overline F_n})$ is jointly Gaussian and $\Cov(\mathcal I(r_n),\mathcal I(u_{n,j}))=\langle r_n,u_{n,j}\rangle_{\Lambda^2}=0$ for every $j$, hence $\mathcal I(r_n)$ is independent of $Y^{\overline F_n}$, and the same holds at the limit.
Writing $a_n=(a_{n,1},\ldots,a_{n,q})^\top$ and $a=(a_1,\ldots,a_q)^\top$, set $X_n\coloneqq a_n^\top Y^{\overline F_n}$, $X\coloneqq a^\top Y^{\overline F}$, $c_n\coloneqq e^{-\norm{r_n}_{\Lambda^2}^2/2}$ and $c\coloneqq e^{-\norm{r}_{\Lambda^2}^2/2}$.
Independence yields
\begin{equation*}
\E\!\left[e^{i\mathcal I(h)}\mid\mathcal A_{\overline F_n}\right]
=\E\!\left[e^{i\mathcal I(r_n)}\mid\mathcal A_{\overline F_n}\right]e^{iX_n}
=\E\!\left[e^{i\mathcal I(r_n)}\right]e^{iX_n}
=c_ne^{iX_n},
\qquad
\E\!\left[e^{i\mathcal I(h)}\mid\mathcal A_{\overline F}\right]
=c e^{iX}.
\end{equation*}
Moreover, $X_n=\mathcal I(h-r_n)\rightarrow\mathcal I(h-r)=X$ in $L^2(\Prob)$, and $\norm{c_ne^{iX_n}-ce^{iX}}_{L^2(\Prob)}\leq\abs{c_n-c}+\norm{X_n-X}_{L^2(\Prob)}\rightarrow0$, where the last inequality uses $\abs{e^{ix}-e^{iy}}\leq\abs{x-y}$ and the isometry of $\mathcal I$.
Taking real and imaginary parts proves the asserted convergence when $Z$ is any finite real linear combination of variables of the form $\cos(\mathcal I(h))$ and $\sin(\mathcal I(h))$.

Let $\mathcal G_M$ be the $\Prob$-completion of $\sigma(M_t:t\in[0,1])$.
Since $M$ has continuous paths, each $M_t^k$ is the almost-sure limit of $M_{t_m}^k$ along any sequence $(t_m)$ in $\mathbb Q\cap[0,1]$ converging to $t$; hence $\mathcal G_M$ is also the $\Prob$-completion of $\sigma(M_t^k:t\in\mathbb Q\cap[0,1],\,k=1,\ldots,d)$.
Moreover, if $e_k$ denotes the $k$th coordinate vector of $\R^d$, then $\mathbf 1_{(0,t]}e_k\in\Lambda^2$ and
\begin{equation*}
\mathcal I(\mathbf 1_{(0,t]}e_k)=\int_0^t e_k^\top\,\mathrm dM_s=M_t^k.
\end{equation*}
It follows that $\mathcal G_M$ is contained in $\sigma(\mathcal I(h):h\in\Lambda^2)$.
For the reverse inclusion, fix $h\in\Lambda^2$ and choose a Borel measurable representative, again denoted by $h$.
Since $\pi(t)$ is positive semidefinite and $\operatorname{tr}\pi(t)=1$ for $\mu$-almost every $t$, its largest eigenvalue is at most one, and hence $v^\top\pi(t)v\leq\norm{v}^2$ for every $v\in\R^d$ and $\mu$-almost every $t$.
For $r\in\mathbb N$, set $h^{(r)}\coloneqq h\mathbf 1_{\{\norm{h}\leq r\}}$.
Then $\norm{h-h^{(r)}}_{\Lambda^2}^2=\int_{\{\norm{h}>r\}} h(t)^\top\pi(t)h(t)\,\mu(\mathrm dt)\rightarrow0$.
Since $\mu$ is finite and $h^{(r)}$ is bounded, $h^{(r)}\in L^2(\mu;\R^d)$.
Step functions are dense in $L^2(\mu;\R^d)$, so for each $r$ we may choose such a function $s_r$ with
$\norm{s_r-h^{(r)}}_{L^2(\mu;\R^d)}\leq r^{-1}$.
The preceding bound on $\pi$ then gives
\begin{equation*}
\norm{s_r-h}_{\Lambda^2}\leq\norm{s_r-h^{(r)}}_{\Lambda^2}+\norm{h^{(r)}-h}_{\Lambda^2}\leq\norm{s_r-h^{(r)}}_{L^2(\mu;\R^d)}+\norm{h^{(r)}-h}_{\Lambda^2}
\rightarrow0.
\end{equation*}
Writing $s_r=\sum_{\ell=1}^{N_r}v_{r,\ell}\mathbf 1_{(t_{r,\ell-1},t_{r,\ell}]}$, we have $\mathcal I(s_r)=\sum_{\ell=1}^{N_r}v_{r,\ell}^\top\bigl(M_{t_{r,\ell}}-M_{t_{r,\ell-1}}\bigr)$, and hence $\mathcal I(s_r)$ is $\mathcal G_M$-measurable.
By the isometry of $\mathcal I$, $\mathcal I(s_r)\to\mathcal I(h)$ in $L^2(\Prob)$.
A subsequence therefore converges almost surely, and completeness of $\mathcal G_M$ implies that $\mathcal I(h)$ is $\mathcal G_M$-measurable.
Consequently, $\mathcal G_M=\sigma(\mathcal I(h):h\in\Lambda^2)$.

Since $\Lambda^2$ is separable, we may choose an orthonormal basis $(\varphi_j)_{j\in J}$, where $J=\{1,\ldots,d_0\}$ if $\Lambda^2$ has finite dimension $d_0$, and $J=\mathbb N$ otherwise, and set $\xi_j\coloneqq\mathcal I(\varphi_j)$.
If $\Lambda^2$ is infinite-dimensional, then for $h_m\coloneqq\sum_{j=1}^m\langle h,\varphi_j\rangle_{\Lambda^2}\varphi_j$ we have $h_m\to h$ in $\Lambda^2$ and therefore
\begin{equation*}
\mathcal I(h_m)=\sum_{j=1}^m\langle h,\varphi_j\rangle_{\Lambda^2}\xi_j\rightarrow\mathcal I(h)\qquad\text{in }L^2(\Prob).
\end{equation*}
In the finite-dimensional case the corresponding representation is an exact finite sum.
Since each $\xi_j$ is itself of the form $\mathcal I(h)$, we conclude in both cases that $\mathcal G_M$ is also the $\Prob$-completion of $\sigma(\xi_j:j\in J)$.

Let $\mathcal E$ be the real linear span of $\{\cos(\mathcal I(h)),\sin(\mathcal I(h)):h\in\Lambda^2\}$.
To prove that $\mathcal E$ is dense in $L^2(\Omega,\mathcal G_M,\Prob)$ (cf. \citet[Remark~2.15]{janson1997gaussian}), let a real $U$ in this space be orthogonal to $\mathcal E$.
For each finite collection of the basis, with $\xi^{(m)}=(\xi_1,\ldots,\xi_m)$, define the finite signed measure $\nu_m(B)\coloneqq\E[U\mathbf 1_{\{\xi^{(m)}\in B\}}]$ on $\R^m$, whose total variation is at most $\E[\abs U]$.
For every $t\in\R^m$, orthogonality to the sine and cosine corresponding to $h=\sum_{j=1}^m t_j\varphi_j$ gives
\begin{equation*}
\widehat\nu_m(t)
=\int_{\R^m}e^{it^\top x}\,\nu_m(\mathrm dx)
=\E[Ue^{it^\top\xi^{(m)}}]=0.
\end{equation*}
Uniqueness of Fourier transforms of finite signed measures yields $\nu_m=0$, and hence $\E[U\mid\sigma(\xi_1,\ldots,\xi_m)]=0$.
In the finite-dimensional case, taking $m=d_0$ gives $U=\E[U\mid\mathcal G_M]=0$; in the infinite-dimensional case, the martingale convergence theorem along the increasing fields $\sigma(\xi_1,\ldots,\xi_m)$ gives $U=0$ in $L^2(\Prob)$.
Thus $\mathcal E$ is dense.

Finally, for arbitrary $Z\in L^2(\Prob)$, set $Z_M\coloneqq\E[Z\mid\mathcal G_M]$.
Since $\mathcal A_{F_n},\mathcal A_F\subseteq\mathcal G_M$, the tower property gives $\E[Z\mid\mathcal A_{F_n}]=\E[Z_M\mid\mathcal A_{F_n}]$ and the analogous identity for $F$.
For $V\in\mathcal E$, contractivity of conditional expectation gives
\begin{equation*}
\norm{\E[Z_M\mid\mathcal A_{F_n}]-\E[Z_M\mid\mathcal A_F]}_{L^2(\Prob)}\leq 2\norm{Z_M-V}_{L^2(\Prob)}+\norm{\E[V\mid\mathcal A_{F_n}]-\E[V\mid\mathcal A_F]}_{L^2(\Prob)}.
\end{equation*}
Given $\varepsilon>0$, choose $V\in\mathcal E$ so that the first term on the right is below $\varepsilon/2$, and then take $n$ large enough that the second is below $\varepsilon/2$.
\end{proof}

\cref{thm:conditional_expectation_stability} yields convergence of conditional expectations, provided that the ranks of the Gram matrices are eventually equal to $\operatorname{rank}\Sigma_F$.
This condition cannot be omitted in general, as the following example shows.
Recall the setting and notation from \cref{ex:martingale_convergence}, and set $F_n\coloneqq(g,\varepsilon_nf)$ and $F\coloneqq(g,0)$.
Because $\varepsilon_n>0$, one has $\mathcal A_{F_n}=\sigma(W,Z)$ and $\mathcal A_F=\sigma(W)$.
Moreover, $\norm{F_n-F}_{(\Lambda^2)^2}=\varepsilon_n\rightarrow0$,
\begin{equation*}
Y^{F_n}=\begin{pmatrix}W\\ \varepsilon_nZ\end{pmatrix}\rightarrow Y^F=\begin{pmatrix}W\\0\end{pmatrix}
\quad\text{in }L^2(\Prob;\R^2),
\end{equation*}
$\operatorname{rank}\Sigma_{F_n}=2$, $\operatorname{rank}\Sigma_F=1$, $\E[G\mid\mathcal A_{F_n}]=Z$, $\E[G\mid\mathcal A_F]=\E[Z\mid\sigma(W)]=0$, and $\norm{\E[G\mid\mathcal A_{F_n}]-\E[G\mid\mathcal A_F]}_{L^2(\Prob)}=1$.
Thus the limiting factor is nonzero and singular, the factors themselves converge in $L^2$, but their conditional expectations converge to $Z \neq 0$.

To pass from conditional expectations to conditional distributions, let $\mathcal P(\R)$ denote the set of Borel probability measures on $\R$, endowed with the weak topology.
The conditional distribution $\mu_Y$ is then a $\mathcal P(\R)$-valued random variable.
For $\varphi\in C_b(\R)$, where $C_b(\R)$ denotes the bounded continuous real-valued functions, put $\operatorname{Lip}_1(\varphi)
\coloneqq\sup_{x\neq y}\frac{\abs{\varphi(x)-\varphi(y)}}{\abs{x-y}\wedge1}$ and
\begin{equation*}
d_{\mathrm{BL}}(\nu,\nu')
\coloneqq\sup\left\{
\abs{\int_\R\varphi\,\mathrm d\nu-\int_\R\varphi\,\mathrm d\nu'}:
\varphi\in C_b(\R),\ \norm{\varphi}_\infty+\operatorname{Lip}_1(\varphi)\leq1
\right\}.
\end{equation*}
The bounded metric $d_{\mathrm{BL}}$ metrises the weak topology on $\mathcal P(\R)$.
Moreover, for each bounded Borel-measurable function $\varphi:\R\to\R$, the defining property of the conditional distribution gives $\int_\R\varphi\,\mathrm d\mu_Y=\E[\varphi(G)\mid\sigma(Y)]$, hence, under the hypotheses of \cref{thm:conditional_expectation_stability}, $\int_\R\varphi\,\mathrm d\mu_{Y^{F_n}}\to\int_\R\varphi\,\mathrm d\mu_{Y^F}$ in $L^2(\Prob)$ for every fixed $\varphi\in C_b(\R)$.
Note that such convergence for each fixed test function does not immediately control the uncountable supremum in $d_{\mathrm{BL}}$.

The following example shows that, at a singular limit, convergence of Gaussian factors does not in general imply convergence of the associated conditional distributions.
In fact, two approximating sequences may have the same covariance matrix at every step and converge to the same limiting factor, while their conditional distributions exhibit different limiting behaviour.
\begin{example}\label{ex:matched_singular_covariance_paths}
Let $d=n=4$ and let $M=(B^1,B^2,B^3,B^4)$ be a four-dimensional standard Brownian motion.
Take $x_0=0$, $a\equiv0$ and $b\equiv I_4$, so that $X=M$, and set $H(x)\coloneqq x_1^{(2)}$.
With $f(t)\equiv e_1$, $g(t)\equiv e_2$, $h(t)\equiv e_3$ and $k(t)\equiv e_4$, write $Z\coloneqq\mathcal I(f)$, $G=\mathcal I(g)$, $U\coloneqq\mathcal I(h)$ and $V\coloneqq\mathcal I(k)$.
By construction, $Z$, $G$, $U$ and $V$ are independent standard normal variables.
Let $\varepsilon_n\coloneqq n^{-1/2}$, put $\ell\coloneqq(g+h)/\sqrt2$, and write $Q\coloneqq\mathcal I(\ell)=(G+U)/\sqrt2$.
We compare the Gaussian factors
\begin{equation*}
Y_n^+=\begin{pmatrix}Z+\varepsilon_nU\\Z+\varepsilon_nV\end{pmatrix},
\qquad
Y_n^-=\begin{pmatrix}Z+\varepsilon_nQ\\Z+\varepsilon_nV\end{pmatrix},
\qquad
Y=\begin{pmatrix}Z\\Z
\end{pmatrix},
\end{equation*}
which are respectively induced by $F_n^+=(f+\varepsilon_nh,f+\varepsilon_nk)$, $F_n^-=(f+\varepsilon_n\ell,f+\varepsilon_nk)$, and
$F=(f,f)$.
Note that $F_n^\pm\to F$ in $(\Lambda^2)^2$, and hence $Y_n^\pm\to Y$ in $L^2(\Prob;\R^2)$.
Moreover, the two approximating factors have the same covariance matrix,
\begin{equation*}
\Sigma_{F_n^+}=\Sigma_{F_n^-}=\begin{pmatrix}1+\varepsilon_n^2 & 1\\1 & 1+\varepsilon_n^2\end{pmatrix}\rightarrow\begin{pmatrix}1 & 1\\1 & 1\end{pmatrix}=\Sigma_F.
\end{equation*}
Both $\Sigma_{F_n^+}$ and $\Sigma_{F_n^-}$ are positive definite, since their determinants
equal $\varepsilon_n^2(2+\varepsilon_n^2)$, whereas $\Sigma_F$ has rank one.
For the first sequence, $G$ is independent of $Y_n^+$ and of $Y$.
Consequently, $m_{Y_n^+}=m_Y=0$ and $\mu_{Y_n^+}=\mu_Y=\mathcal N(0,1)$ for every $n$.
We conclude that the asymptotic loss of rank for $\Sigma_{F_n^+}$ does not in itself prevent convergence of the conditional distributions $\mu_{Y_n^+}$ to $\mu_Y$.

On the other hand, since $\Cov(G,Y_n^-)=(\varepsilon_n/\sqrt2,0)$, direct computation gives
\begin{equation*}
m_{Y_n^-}=\frac{(1+\varepsilon_n^2)Y_{n,1}^- -Y_{n,2}^-}{\sqrt2\,\varepsilon_n(2+\varepsilon_n^2)}\rightarrow\frac{Q-V}{2\sqrt2}\quad\text{in }L^2(\Prob),
\qquad
\Var(G\mid\sigma(Y_n^-))=\frac{3+\varepsilon_n^2}{4+2\varepsilon_n^2}\rightarrow\frac34,
\end{equation*}
where the first convergence follows upon substituting $Y_{n,1}^-=Z+\varepsilon_nQ$ and $Y_{n,2}^-=Z+\varepsilon_nV$.
Writing $m_n\coloneqq m_{Y_n^-}$ and $s_n^2\coloneqq(3+\varepsilon_n^2)/(4+2\varepsilon_n^2)$, we therefore have $\mu_{Y_n^-}=\mathcal N(m_n,s_n^2)$, whereas $\mu_Y=\mathcal N(0,1)$.
We will now show that these conditional distributions remain uniformly separated in the $d_{\mathrm{BL}}$ metric.
To this end, note that the functions $x\mapsto\cos(x)/3$ and $x\mapsto\sin(x)/3$ are admissible test functions for the $d_{\mathrm{BL}}$ metric, since their supremum norms equal $1/3$ and their $\operatorname{Lip}_1$-seminorms are at most $2/3$.
Put $\Delta_n\coloneqq \int_\R \exp(i x)(\mu_{Y_n^-} - \mu_Y)(\mathrm{d}x) = \exp(i m_n-s_n^2/2)-\exp(-1/2)$.
Using the characteristic function for the Gaussian distribution, the reverse triangle inequality, and the fact that $\max\{\abs{a},\abs{b}\}\geq\abs{z}/\sqrt2$ for every $z = a + i b\in\mathbb C$ (in particular for $z=\Delta_n$), we have
\begin{equation*}
d_{\mathrm{BL}}(\mu_{Y_n^-},\mu_Y)
\geq
\frac{\abs{\Delta_n}}{3\sqrt2}
\geq
\frac{\exp(-s_n^2/2)-\exp(-1/2)}{3\sqrt2}
\geq
\frac{\exp(-3/8)-\exp(-1/2)}{3\sqrt2}
\eqqcolon c_0>0.
\end{equation*}
We thus conclude that $\Prob(d_{\mathrm{BL}}(\mu_{Y_n^-},\mu_Y)\geq c_0)=1$ for every $n$, and $\mu_{Y_n^-}$ does not converge to $\mu_Y$ in probability.
\end{example}

We recall the following standard construction of a metric for weak convergence, which will be useful in the proof of \cref{thm:conditional_distribution_stability}; see \citet[Chapter~3, Proposition~4.4]{ethier86markov} for the convergence-determining property of compactly supported continuous functions.
\begin{lemma}\label{lem:weak_convergence_determining_class}
There exists a sequence $(\psi_j)_{j\geq1}\subset C_b(\R)$ with
$\norm{\psi_j}_\infty\leq1$ such that, for every sequence
$(\nu_n)_{n\geq1}\subset\mathcal P(\R)$ and every $\nu\in\mathcal P(\R)$,
\begin{equation*}
\nu_n\Rightarrow\nu\quad\Longleftrightarrow\quad\int_\R\psi_j\,\mathrm d\nu_n\rightarrow\int_\R\psi_j\,\mathrm d\nu\quad\text{for every }j\geq1.
\end{equation*}
For this sequence,
\begin{equation*}
d_*(\nu,\nu')\coloneqq\sum_{j=1}^\infty 2^{-j}\abs{\int_\R\psi_j\,\mathrm d\nu-\int_\R\psi_j\,\mathrm d\nu'},\qquad \nu,\nu'\in\mathcal P(\R),
\end{equation*}
defines a metric, bounded above by $2$, which metrises the weak topology on $\mathcal P(\R)$.
\end{lemma}

\begin{proof}
To prove the equivalence, it is enough to show that convergence of the test integrals for a suitable countable family of bounded continuous functions implies weak convergence, as the converse implication is immediate from the definition of weak convergence.
Let $C_0(\R)$ denote the Banach space of continuous real-valued functions on $\R$ which vanish at infinity, equipped with the uniform norm.
Since $C_0(\R)$ is separable, its closed unit ball contains a countable dense subset, which we enumerate as $(\psi_j)_{j\geq1}$.
Now suppose that the test integrals $\int_\R\psi_j\,\mathrm d\nu_n$ converge to $\int_\R\psi_j\,\mathrm d\nu$ for every $j\geq1$.
For $0\neq h\in C_0(\R)$ and $j\geq1$,
\begin{equation*}
\abs{\int_\R h\,\mathrm d\nu_n-\int_\R h\,\mathrm d\nu}\leq \norm{h}_\infty\abs{\int_\R\psi_j\,\mathrm d\nu_n-\int_\R\psi_j\,\mathrm d\nu}+2\norm{h-\norm{h}_\infty\psi_j}_\infty,
\end{equation*}
where the first term on the right tends to zero as $n\to\infty$, and the second term can be made arbitrarily small by density of the $\psi_j$.
The case $0=h\in C_0(\R)$ is trivial.

What remains is to extend convergence of the test integrals from $C_0(\R)$ to $C_b(\R)$.
To this end, fix $f\in C_b(\R)$ and $\varepsilon\in(0,1)$.
Choose $R>0$ such that $\nu([-R,R])>1-\varepsilon$, and choose a continuous function $\chi:\R\to[0,1]$ with compact support and $\chi=1$ on $[-R,R]$.
Now $\int_\R(1-\chi)\,\mathrm d\nu<\varepsilon$, and since $\chi\in C_0(\R)$, there exists $n_0\in\mathbb N$ such that $\int_\R(1-\chi)\,\mathrm d\nu_n<2\varepsilon$ for all $n\geq n_0$.
Moreover, $f = f\chi + f(1-\chi)$, and $f\chi\in C_0(\R)$, hence
\begin{equation*}
\abs{\int_\R f\,\mathrm d\nu_n-\int_\R f\,\mathrm d\nu}\leq\abs{\int_\R f\chi\,\mathrm d\nu_n-\int_\R f\chi\,\mathrm d\nu}+\norm{f}_\infty\left(\int_\R(1-\chi)\,\mathrm d\nu_n+\int_\R(1-\chi)\,\mathrm d\nu\right),
\end{equation*}
where the first term on the right tends to zero as $n\to\infty$, and the other two are eventually bounded by $3\norm{f}_\infty\varepsilon$.
Letting $\varepsilon\downarrow0$ proves $\nu_n\Rightarrow\nu$.

In order to see that $d_*$ is a metric, note that the $\psi_j$ are bounded by $1$ in supremum norm, so each difference of test integrals is at most $2$ and thus $d_*(\nu,\nu')\leq2$.
Symmetry and the triangle inequality are also immediate from the definition of $d_*$.
If $d_*(\nu,\nu')=0$, then $\int_\R\psi_j\,\mathrm d\nu=\int_\R\psi_j\,\mathrm d\nu'$ for every $j\geq1$.
For $0\neq h\in C_0(\R)$ and $j\geq1$,
\begin{equation*}
\abs{\int_\R h\,\mathrm d\nu-\int_\R h\,\mathrm d\nu'}\leq2\norm{h-\norm{h}_\infty\psi_j}_\infty,
\end{equation*}
where the right-hand side can be made arbitrarily small.
Thus $\int_\R h\,\mathrm d\nu=\int_\R h\,\mathrm d\nu'$ for every $h\in C_0(\R)$, and therefore $\nu=\nu'$, since probability measures on $\R$ are determined by their integrals against $C_0(\R)$ by the Riesz--Markov--Kakutani representation theorem.
We conclude that $d_*$ is a metric.

In order to see that $d_*$ metrises the weak topology, note that $d_{\mathrm{BL}}$ metrises the weak topology, and that two metrizable topologies on the same set coincide if they have the same convergent sequences with the same limits.
If $\nu_n\Rightarrow\nu$, then $\int_\R\psi_j\,\mathrm d\nu_n\to\int_\R\psi_j\,\mathrm d\nu$ for every $j\geq1$.
Moreover, since $\norm{\psi_j}_\infty\leq1$, $2^{-j}\abs{\int_\R\psi_j\,\mathrm d\nu_n-\int_\R\psi_j\,\mathrm d\nu}\leq2^{1-j}$, and $\sum_{j\geq1}2^{1-j}<\infty$.
Dominated convergence for series therefore yields $d_*(\nu_n,\nu)\to0$.
Conversely, $d_*(\nu_n,\nu)\to0$ implies, for every $j\geq1$,
\begin{equation*}
\abs{\int_\R\psi_j\,\mathrm d\nu_n-\int_\R\psi_j\,\mathrm d\nu}\leq 2^j d_*(\nu_n,\nu)\rightarrow0,
\end{equation*}
and the first part of the proof gives $\nu_n\Rightarrow\nu$.
Thus $d_*(\nu_n,\nu)\to0$ if and only if $\nu_n\Rightarrow\nu$.
Since $d_{\mathrm{BL}}$ metrises the weak topology on $\mathcal P(\R)$, the metrics $d_*$ and $d_{\mathrm{BL}}$ induce the same topology.
\end{proof}

\begin{theorem}\label{thm:conditional_distribution_stability}
Let $q\in\mathbb N$ and let $F_n,F\in(\Lambda^2)^q$ satisfy $F_n\to F$ in $(\Lambda^2)^q$ and $\operatorname{rank}\Sigma_{F_n}=\operatorname{rank}\Sigma_F$ for all sufficiently large $n$.
Then, for every $p\in[1,\infty)$,
\begin{equation*}
d_{\mathrm{BL}}\bigl(\mu_{Y^{F_n}},\mu_{Y^F}\bigr)\rightarrow0\qquad\text{in }L^p(\Prob).
\end{equation*}
In particular, $\mu_{Y^{F_n}}$ converges to $\mu_{Y^F}$ in probability in the weak topology on $\mathcal P(\R)$.
\end{theorem}

\begin{proof}
Fix a sequence $(\psi_j)_{j\geq1}$ and the associated metric $d_*$ furnished by \cref{lem:weak_convergence_determining_class}, and recall that $d_*$ metrises the weak topology on $\mathcal P(\R)$.
For $n,j\geq1$, put
\begin{equation*}
U_{n,j}\coloneqq\abs{\int_\R\psi_j\,\mathrm d\mu_{Y^{F_n}}-\int_\R\psi_j\,\mathrm d\mu_{Y^F}},
\qquad
D_n^*\coloneqq d_*\bigl(\mu_{Y^{F_n}},\mu_{Y^F}\bigr),
\quad
D_n^{\mathrm{BL}}\coloneqq d_{\mathrm{BL}}\bigl(\mu_{Y^{F_n}},\mu_{Y^F}\bigr).
\end{equation*}
Since $\psi_j\in C_b(\R)$, the map $\nu\mapsto\int_\R\psi_j\,\mathrm d\nu$ is continuous for the weak topology.
Since $\mu_{Y^{F_n}}$ and $\mu_{Y^F}$ are Borel measurable, both $\int_\R\psi_j\,\mathrm d\mu_{Y^{F_n}}$ and
$\int_\R\psi_j\,\mathrm d\mu_{Y^F}$ are measurable.
Hence $U_{n,j}$ is measurable.
Moreover, $D_n^*$ is measurable as the pointwise limit of its measurable partial sums.
The continuity of $d_{\mathrm{BL}}$ for the product weak topology also shows that $D_n^{\mathrm{BL}}$ is measurable.
For each $n,j\geq1$, the defining properties of the random conditional distributions give
\begin{equation*}
\int_\R\psi_j\,\mathrm d\mu_{Y^{F_n}}=\E[\psi_j(G)\mid\mathcal A_{F_n}],
\qquad
\int_\R\psi_j\,\mathrm d\mu_{Y^F}=\E[\psi_j(G)\mid\mathcal A_F].
\end{equation*}
Since $\norm{\psi_j}_\infty\leq1$, one has $\psi_j(G)\in L^2(\Prob)$ and $0\leq U_{n,j}\leq2$.
Thus \cref{thm:conditional_expectation_stability} yields $\norm{U_{n,j}}_{L^2(\Prob)}\to0$ for every $j\geq1$.

For fixed $n$, let $S_{n,m}\coloneqq\sum_{j=1}^m2^{-j}U_{n,j}$.
Since $0\leq D_n^*-S_{n,m}\leq2\sum_{j=m+1}^\infty2^{-j}=2^{1-m}$, one has $S_{n,m}\to D_n^*$ in $L^2(\Prob)$.
For each $m\in\mathbb N$, Minkowski's inequality gives $\norm{S_{n,m}}_{L^2(\Prob)}\leq\sum_{j=1}^m2^{-j}\norm{U_{n,j}}_{L^2(\Prob)}$.
Since $S_{n,m}\to D_n^*$ in $L^2(\Prob)$, letting $m\to\infty$ yields
\begin{equation*}
\norm{D_n^*}_{L^2(\Prob)}
\leq
\sum_{j=1}^\infty
2^{-j}\norm{U_{n,j}}_{L^2(\Prob)}.
\end{equation*}
Each summand on the right tends to zero and is bounded by $2^{1-j}$.
Dominated convergence for series hence shows that $D_n^*\to0$ in $L^2(\Prob)$.

For every subsequence $(n_k)_{k\geq1}$, there is a further subsequence $(n_{k_\ell})_{\ell\geq1}$ such that $D_{n_{k_\ell}}^*\to0$ almost surely.
For every $\omega$ outside the corresponding null set, \cref{lem:weak_convergence_determining_class} implies that $\mu_{Y^{F_{n_{k_\ell}}}}(\omega)$ converges weakly to $\mu_{Y^F}(\omega)$, and hence $D_{n_{k_\ell}}^{\mathrm{BL}}(\omega)\to0$ because $d_{\mathrm{BL}}$ metrises the weak topology.
If $D_n^{\mathrm{BL}}$ did not converge to zero in probability, some subsequence would admit $\varepsilon,\delta>0$ with $\Prob(D_{n_k}^{\mathrm{BL}}>\varepsilon)\geq\delta$ for every $k$.
But for that subsequence, one could then choose a further subsequence along which $D_{n_{k_\ell}}^{\mathrm{BL}}\to0$ almost surely, which yields a contradiction.
Thus $D_n^{\mathrm{BL}}\to0$ in probability.
Finally, $0\leq D_n^*,D_n^{\mathrm{BL}}\leq2$.
For either sequence, denoted by $(D_n)$, every $p\in[1,\infty)$ and every $\varepsilon>0$ satisfy
\begin{equation*}
\E[D_n^p]\leq\varepsilon^p+2^p\Prob(D_n>\varepsilon).
\end{equation*}
Since both sequences converge to zero in probability, first letting $n\to\infty$ and then $\varepsilon\downarrow0$ proves the asserted $L^p(\Prob)$ convergences.
\end{proof}

After having established stability of conditional expectations and conditional distributions, we now show that, for every $F\in(\Lambda^2)^q$ with $\Sigma_F$ positive definite, neural stratification factors approximate $m_{Y^F}$ and $\mu_{Y^F}$ arbitrarily well.
To this end, fix a bounded, continuous and nonconstant activation function $\psi:\R\to\R$.
Let $\mathfrak N_\psi^d$ denote the set of all functions of the form
\begin{equation*}
[0,1]\ni t\mapsto b+\sum_{\ell=1}^m v_\ell\psi(\alpha_\ell t+\beta_\ell),
\qquad
m\in\mathbb N,\quad b,v_1,\ldots,v_m\in\R^d,\quad \alpha_1,\ldots,\alpha_m,\beta_1,\ldots,\beta_m\in\R.
\end{equation*}
Every element of $\mathfrak N_\psi^d$ is bounded and continuous and hence represents an element of $\Lambda^2$.
By \citet[Proposition~3.4 and Lemma~2.10(c)]{arandjelovic25importance}, $\mathfrak N_\psi^d$ is dense in $\Lambda^2$.
We henceforth take the theoretical parameter set $\Theta$ from \cref{eq:optimization_problem} to index the full product $(\mathfrak N_\psi^d)^q$, so that $\{F_\theta:\theta\in\Theta\}=(\mathfrak N_\psi^d)^q$.

\begin{corollary}\label{cor:neural_stratification_approximation}
Let $F\in(\Lambda^2)^q$ be such that $\Sigma_F$ is positive definite.
Then there exists a sequence $(\theta_n)_{n\in\mathbb N}$ in $\Theta_\mathrm{nd}$ such that $F_{\theta_n}\to F$ in $(\Lambda^2)^q$, $m_{Y_{\theta_n}}\to m_{Y^F}$ in $L^2(\Prob)$, and $d_{\mathrm{BL}}(\mu_{Y_{\theta_n}},\mu_{Y^F})\to0$ in $L^p(\Prob)$ for every $p\in[1,\infty)$.
Moreover, $\Var(m_{Y_{\theta_n}})\to\Var(m_{Y^F})$, and
\begin{equation*}
\sup_{\theta\in\Theta_\mathrm{nd}}\Var(m_{Y_\theta})=\sup_{\substack{\widetilde F\in(\Lambda^2)^q\\ \Sigma_{\widetilde F}\text{ positive definite}}} \Var(m_{Y^{\widetilde F}}).
\end{equation*}
In particular, for every $\varepsilon>0$ there exists $\theta\in\Theta_\mathrm{nd}$ such that $\Var(m_{Y_\theta})\geq\Var(m_{Y^F})-\varepsilon$.
\end{corollary}

\begin{proof}
Write $F=(f_1,\ldots,f_q)$.
For each $n\in\mathbb N$ and $j=1,\ldots,q$, we may choose $g_{n,j}\in\mathfrak N_\psi^d$ such that $\norm{g_{n,j}-f_j}_{\Lambda^2}<1/(n\sqrt q)$.
Since $\Theta$ indexes $(\mathfrak N_\psi^d)^q$, there is $\theta_n\in\Theta$ with $F_{\theta_n}=(g_{n,1},\ldots,g_{n,q})$, and
\begin{equation*}
\norm{F_{\theta_n}-F}_{(\Lambda^2)^q}^2 = \sum_{j=1}^q\norm{g_{n,j}-f_j}_{\Lambda^2}^2 <\frac1{n^2}.
\end{equation*}
Thus $F_{\theta_n}\to F$ in $(\Lambda^2)^q$.

For $i,j=1,\ldots,q$, the Cauchy--Schwarz inequality gives
\begin{equation*}
\abs{\langle g_{n,i},g_{n,j}\rangle_{\Lambda^2}-\langle f_i,f_j\rangle_{\Lambda^2}}\leq\norm{g_{n,i}-f_i}_{\Lambda^2}\norm{g_{n,j}}_{\Lambda^2}+\norm{f_i}_{\Lambda^2}\norm{g_{n,j}-f_j}_{\Lambda^2}\rightarrow0.
\end{equation*}
Hence $\Sigma_{F_{\theta_n}}\to\Sigma_F$ entrywise and therefore in operator norm (since the operator norm is dominated by the Frobenius norm).
Since $\lambda_{\min}(\Sigma_F)>0$, continuity of the smallest eigenvalue on the space of symmetric matrices, which follows from Weyl's inequality, implies that $\Sigma_{F_{\theta_n}}$ is positive definite for all sufficiently large $n$.
After discarding finitely many terms and relabelling, we have $\theta_n\in\Theta_\mathrm{nd}$ for every $n$.

For such $n$, the matrix $\Sigma_{F_{\theta_n}}^{-1/2}$ is deterministic and invertible, and the definition of $Y_{\theta_n}$ gives $Y_{\theta_n}=\Sigma_{F_{\theta_n}}^{-1/2}Y^{F_{\theta_n}}$.
Consequently, $\sigma(Y_{\theta_n})=\sigma(Y^{F_{\theta_n}}) =\mathcal A_{F_{\theta_n}}$.
Both $\Sigma_{F_{\theta_n}}$ and $\Sigma_F$ have rank $q$, and hence \cref{thm:conditional_expectation_stability,thm:conditional_distribution_stability} yield
\begin{equation*}
m_{Y_{\theta_n}}\rightarrow m_{Y^F}\ \text{in }L^2(\Prob)
\qquad\text{and}\qquad
d_{\mathrm{BL}}\bigl(\mu_{Y_{\theta_n}},\mu_{Y^F}\bigr)\rightarrow0
\ \text{in }L^p(\Prob),\quad p\in[1,\infty).
\end{equation*}
Finally, since $\E[m_{Y_{\theta_n}}]=\E[m_{Y^F}]=\E[G]$, the $L^2$-convergence of the conditional means implies $\Var(m_{Y_{\theta_n}})\to\Var(m_{Y^F})$.

Set $J_{\mathrm{NN}}^\ast\coloneqq\sup_{\theta\in\Theta_\mathrm{nd}}\Var(m_{Y_\theta})$ and $J^\ast\coloneqq\sup\{\Var(m_{Y^F}):F\in(\Lambda^2)^q,\ \Sigma_F\text{ positive definite}\}$.
Since $F_\theta$ is admissible in the definition of $J^\ast$ for every $\theta\in\Theta_\mathrm{nd}$, we have $J_{\mathrm{NN}}^\ast\leq J^\ast$.
Conversely, let $\delta>0$ and choose $\widetilde F\in(\Lambda^2)^q$ with $\Sigma_{\widetilde F}$ positive definite and $\Var(m_{Y^{\widetilde F}})\geq J^\ast-\delta$.
For every $\varepsilon>0$, there exists $\theta\in\Theta_\mathrm{nd}$ such that
\begin{equation*}
J_{\mathrm{NN}}^\ast\geq\Var(m_{Y_\theta})\geq\Var(m_{Y^{\widetilde F}})-\varepsilon\geq J^\ast-\delta-\varepsilon.
\end{equation*}
Letting first $\varepsilon\downarrow0$ and then $\delta\downarrow0$ gives $J_{\mathrm{NN}}^\ast\geq J^\ast$, and hence $J_{\mathrm{NN}}^\ast=J^\ast$.
\end{proof}

\section{Numerical applications to derivative pricing}\label{sec:simulation_study}
We now examine the variance-reduction performance of neural stratification factors for a dual-condition CMS range accrual in a G2++ model and a worst-of autocallable in a Wishart stochastic-covariance model.
We use the $L^2$-projection identity
\begin{equation*}
\E[\Var(G\mid\sigma(Y))]=\inf_{h\in L^2(\mathcal L(Y))}\E[(G-h(Y))^2]
\end{equation*}
to jointly minimise $\E[(G-h_\vartheta(Y_\theta))^2]$ over the parameters of the factor network and a feedforward neural network $h_\vartheta$.
The neural parametrisation is convenient because it allows refinement of the time discretisation without increasing the number of parameters.
For $q\in\{1,2,3\}$, each factor coordinate is divided into $b_q\in\mathbb N$ equiprobable standard-normal quantile intervals, where we take $b_1=4096$, $b_2=64$, and $b_3=16$, and the strata are obtained by taking the corresponding Cartesian products (in particular, $b_q^q=4096$ for every $q$).
Conditional samples are generated by Gaussian conditioning of the discretised driving increments.
Each estimator uses $N=2^{19}=524\,288$ realisations of the payoff, so proportional allocation assigns $128$ draws to every stratum.
Two of the most important metrics for evaluating a stratification factor are the explained-variance ratio and the variance-reduction factor.
The former is estimated as a bias-corrected nested Monte Carlo estimate of $\Var(\E[G\mid\sigma(Y)])/\Var(G)$ based on $4096$ draws of $Y$ and $256$ draws of the corresponding conditional payoffs, while the latter is defined as the ratio of the estimated variance of crude Monte Carlo to that of the corresponding stratified estimator.

As benchmark factors, we adapt the two heuristic approaches formulated in \citet{glasserman99asymptotically}.
For either of the two examples considered below, write $\xi\sim\mathcal N(0,I_n)$ for the standardized vector of time-discretized increments and $\widetilde G:\R^n\to(0,\infty)$ for the positive payoff or, if the payoff is not smooth, a smooth proxy thereof.
We first compute $z^\ast$ by numerically maximising the objective $z\mapsto\log\widetilde G(z)-\frac12 z^\top z$, with $z$ parametrised by a neural network.
The first benchmark factor, which we denote \emph{GHS drift}, is then given by $(z^\ast)^\top\xi/\norm{z^\ast}$, and we note that this factor is only available for $q=1$.
The second heuristic considers the eigenpairs $(\lambda_i,v_i)_{i=1}^n$ of $\nabla^2\log\widetilde G(z^\ast)$, with the $v_i$ orthonormal and indexed so that
\begin{equation*}
\left(\frac{\lambda_1}{1-\lambda_1}\right)^2 \geq\cdots\geq \left(\frac{\lambda_n}{1-\lambda_n}\right)^2.
\end{equation*}
For dimension $q\in\{1,2,3\}$, the second benchmark factor, which we denote \emph{GHS Hessian}, is then given by $(v_1^\top\xi,\ldots,v_q^\top\xi)^\top$.

\subsection{CMS range accrual in the G2++ model}\label{subsec:cms-range-accrual}
Our first example is a dual-condition constant-maturity swap (CMS) range accrual in the G2++ model \citep[Section~4.2]{brigo06interest}, where, under a pricing measure $\Prob$, the short rate is given by $r_t=\varphi(t)+x_t+y_t$, with
\begin{equation*}
\mathrm dx_t=-a x_t\,\mathrm dt+\sigma\,\mathrm dW_t^1,\qquad\mathrm dy_t=-b y_t\,\mathrm dt+\eta\bigl(\rho\,\mathrm dW_t^1+\sqrt{1-\rho^2}\,\mathrm dW_t^2\bigr),
\end{equation*}
for independent standard Brownian motions $W^1$ and $W^2$.
We take $x_0=y_0=0$, $(a,\sigma,b,\eta,\rho)=(0.10,0.010,0.50,0.015,-0.70)$, and a five-year horizon.
For $\kappa>0$ and $u\geq0$, set
\begin{equation*}
B_\kappa(u)\coloneqq\frac{1-\exp(-\kappa u)}{\kappa},\qquad\mathcal V(u)\coloneqq\int_0^u\bigl(\sigma^2B_a(s)^2+\eta^2B_b(s)^2+2\rho\sigma\eta B_a(s)B_b(s)\bigr)\,\mathrm ds.
\end{equation*}
For the flat initial discount curve $P(0,u)=\exp(-0.02u)$, the deterministic shift and zero-coupon bond prices can explicitly be expressed as
\begin{equation*}
\begin{aligned}
\varphi(t)
&=0.02+\frac12\bigl(\sigma^2B_a(t)^2+\eta^2B_b(t)^2+2\rho\sigma\eta B_a(t)B_b(t)\bigr),\\
P(t,t+u)
&=\exp\left(-0.02u+\frac12\bigl(\mathcal V(u)-\mathcal V(t+u)+\mathcal V(t)\bigr)-B_a(u)x_t-B_b(u)y_t\right).
\end{aligned}
\end{equation*}
We simulate the model on the business-day grid $t_k=k\Delta$, $k=0,\ldots,1260$, with $\Delta=1/252$, write $x_k$ and $y_k$ for the numerical states at $t_k$, and let $\xi_{k,j}$, $k=1,\ldots,1260$ and $j=1,2$, be independent standard normal random variables.
Writing $r_k\coloneqq\varphi(t_k)+x_k+y_k$, the discount factors are approximated by $\delta_0=1$ and $\delta_{k+1}=\delta_k\exp(-\frac12(r_k+r_{k+1})\Delta)$.

For $m\in\{2,10\}$, let $R_t^{(m)}$ denote the $m$-year CMS rate with semiannual fixed payments,
\begin{equation*}
R_t^{(m)}\coloneqq
\frac{1-P(t,t+m)}{\frac12\sum_{\ell=1}^{2m}P(t,t+\ell/2)},
\end{equation*}
and set $S_t\coloneqq R_t^{(10)}-R_t^{(2)}$.
$R_t^{(m)}$ is the par fixed rate of an $m$-year swap initiated at time $t$, i.e., the rate for which the fixed and floating legs have the same value at inception. The spread $S_t$ measures the slope between the ten- and two-year points of the swap curve, with $S_t>0$ corresponding to an upward-sloping and $S_t<0$ to an inverted curve between these maturities.
We consider a five-year contract with annual coupon rate $c=0.06$, paid quarterly, whose accrual depends jointly on $R^{(10)}$ and $S$.
At each monitoring date $t_k$, the coupon accrues only when both $R_{t_k}^{(10)}$ and $S_{t_k}$ lie within prescribed corridors around their initial values.
With corridor half-widths $h_R=0.008$ and $h_S=0.004$, let
\begin{equation*}
C_k\coloneqq\bigl\{\abs{R_{t_k}^{(10)}-R_0^{(10)}}\leq h_R,\ \abs{S_{t_k}-S_0}\leq h_S\bigr\},\qquad k=1,\ldots,1260,
\end{equation*}
and write $I_k\coloneqq\mathbf 1_{C_k}$.
There are $63$ monitoring dates per quarter, and the corresponding accrual fraction is $A_j\coloneqq 63^{-1}\sum_{k=63(j-1)+1}^{63j}I_k$, $j=1,\ldots,20$, with payment at $t_{63j}=j/4$.
Assuming unit notional, the discounted payoff is $G\coloneqq(c/4)\sum_{j=1}^{20}\delta_{63j}A_j$.
Since the corridor indicators $I_k$ make $G$ discontinuous as a function of $\xi$, we use a smooth proxy to construct derivative-based benchmarks (in particular, the GHS benchmarks).
For $h,\varepsilon>0$, we define
\begin{equation*}
w_{h,\varepsilon}(z)\coloneqq\frac{1}{1+\exp(-(z+h)/\varepsilon)}\frac{1}{1+\exp(-(h-z)/\varepsilon)}.
\end{equation*}
Replacing $I_k$ by $\widetilde I_k\coloneqq w_{h_R,h_R/10}(R_{t_k}^{(10)}-R_0^{(10)})w_{h_S,h_S/10}(S_{t_k}-S_0)$ then yields the proxy $\widetilde G$.

Structured products such as CMS range-accrual notes are issued by major banks and listed for trading on securities exchanges.
For example, in July 2024, NatWest Markets N.V.\ CMS range-accrual notes with an aggregate nominal amount of EUR~25 million and maturity in 2031 were listed on Euronext Dublin's Global Exchange Market (ISIN XS2855989153).
In March 2021, the Bank of Nova Scotia issued USD~50 million of callable range-accrual notes linked jointly to the ten-year CMS rate and the thirty-year minus two-year CMS spread (ISIN XS2320745230).
In August 2018, DBS Bank issued USD~30 million of callable CMS range-accrual notes whose accrual factor is the fraction of days on which both a lower bound on a CMS reference spread and a corridor condition on a CMS reference rate are satisfied (ISIN XS1865518267).
The latter two notes are structurally close to our example in the sense that they incorporate joint conditions involving a CMS rate and a CMS spread.

The structure of the payoff motivates two further benchmark factors.
Define
\begin{equation*}
\overline R^{(10)}(\xi)\coloneqq\frac{1}{1260}\sum_{k=1}^{1260}R_{t_k}^{(10)}(\xi),
\qquad
\overline S(\xi)\coloneqq\frac{1}{1260}\sum_{k=1}^{1260}S_{t_k}(\xi),
\end{equation*}
and let $u_R\coloneqq\nabla\overline R^{(10)}(0)/\norm{\nabla\overline R^{(10)}(0)}$ and $u_S\coloneqq\nabla\overline S(0)/\norm{\nabla\overline S(0)}$.
We use $u_R^\top\xi$ and $u_S^\top\xi$ as the one-dimensional \emph{level} and \emph{slope} factors.
The vectors $u_R$ and $u_S$ represent the directions of the Gaussian shocks to which the average 10-year CMS rate and the average CMS spread are locally most sensitive, respectively; they are combined into the two-dimensional \emph{level--slope} factor when $q=2$.
Using the smooth surrogate $\widetilde G$, we also include the one-dimensional \emph{smooth-gradient} factor $\nabla\widetilde G(0)^\top\xi/\norm{\nabla\widetilde G(0)}$, which represents the direction of the Gaussian shocks to which the smooth payoff is locally most sensitive.

\begin{table}[tbp]
\centering
\begin{threeparttable}
\caption{Stratification results for a CMS range accrual in the G2++ model.}
\label{tab:gauss2pp-stratification}
\begin{tabular}{@{}ccccccc@{}}
\toprule
q & Method & Bins & Estimate & SE & EV ratio & VR \\
\midrule
0 & crude MC & \textemdash{} & 0.14852816 & $8.561\times 10^{-5}$ & \textemdash{} & 1.000 \\
\midrule
1 & level & 4096 & 0.14853275 & $6.376\times 10^{-5}$ & 0.444 & 1.803 \\
1 & slope & 4096 & 0.14855131 & $7.255\times 10^{-5}$ & 0.276 & 1.393 \\
1 & smooth-gradient & 4096 & 0.14837178 & $6.989\times 10^{-5}$ & 0.340 & 1.501 \\
1 & GHS drift & 4096 & 0.14846333 & $7.557\times 10^{-5}$ & 0.229 & 1.283 \\
1 & GHS Hessian & 4096 & 0.14854130 & $6.365\times 10^{-5}$ & 0.454 & 1.809 \\
1 & \textbf{learned} & 4096 & 0.14855629 & $6.207\times 10^{-5}$ & \textbf{0.486} & \textbf{1.903} \\
\midrule
2 & level--slope & $64\times64$ & 0.14849012 & $4.427\times 10^{-5}$ & 0.736 & 3.740 \\
2 & GHS Hessian & $64\times64$ & 0.14856333 & $4.395\times 10^{-5}$ & 0.742 & 3.794 \\
2 & \textbf{learned} & $64\times64$ & 0.14850027 & $4.111\times 10^{-5}$ & \textbf{0.762} & \textbf{4.337} \\
\midrule
3 & GHS Hessian & $16\times16\times16$ & 0.14852578 & $3.857\times 10^{-5}$ & 0.811 & 4.926 \\
3 & \textbf{learned} & $16\times16\times16$ & 0.14852668 & $3.668\times 10^{-5}$ & \textbf{0.879} & \textbf{5.448} \\
\bottomrule
\end{tabular}
\begin{tablenotes}[flushleft]
\footnotesize
\setlength{\leftmargin}{0pt}
\setlength{\itemindent}{0pt}
\setlength{\labelsep}{0pt}
\item[]$q$ denotes the factor dimension. SE: standard error; EV ratio: $\Var(\E[G\mid\sigma(Y)])/\Var(G)$; VR: variance reduction relative to crude Monte Carlo.
\end{tablenotes}
\end{threeparttable}
\end{table}

\Cref{tab:gauss2pp-stratification} reports a crude Monte Carlo estimate of $0.14853$.
The stratified estimates remain close to this value relative to their standard errors, while every factor reduces the estimated variance.
For $q=1$, the level and GHS Hessian factors perform almost identically, with estimated explained-variance (EV) ratios near $45\%$, while slope, smooth-gradient, and GHS drift factors are less effective.
Our learned factor yields an estimated EV ratio of about $49\%$ and a variance-reduction (VR) factor of about $1.9$.
Passing to $q=2$ produces a larger gain.
Both the level--slope and GHS Hessian factors have estimated EV ratios of about $74\%$ and VR factors of $3.7$ and $3.8$, respectively, while our learned factor raises these values to about $76\%$ and $4.3$.
For $q=3$, our learned factor has an estimated EV ratio of about $88\%$ and a VR factor above $5$, compared with about $81\%$ and $4.9$ for the GHS Hessian factor.
We conclude that our learned factor empirically outperforms all considered benchmarks across all factor dimensions.

\begin{figure}[tbp]
\centering
\includegraphics[width=.9\linewidth]{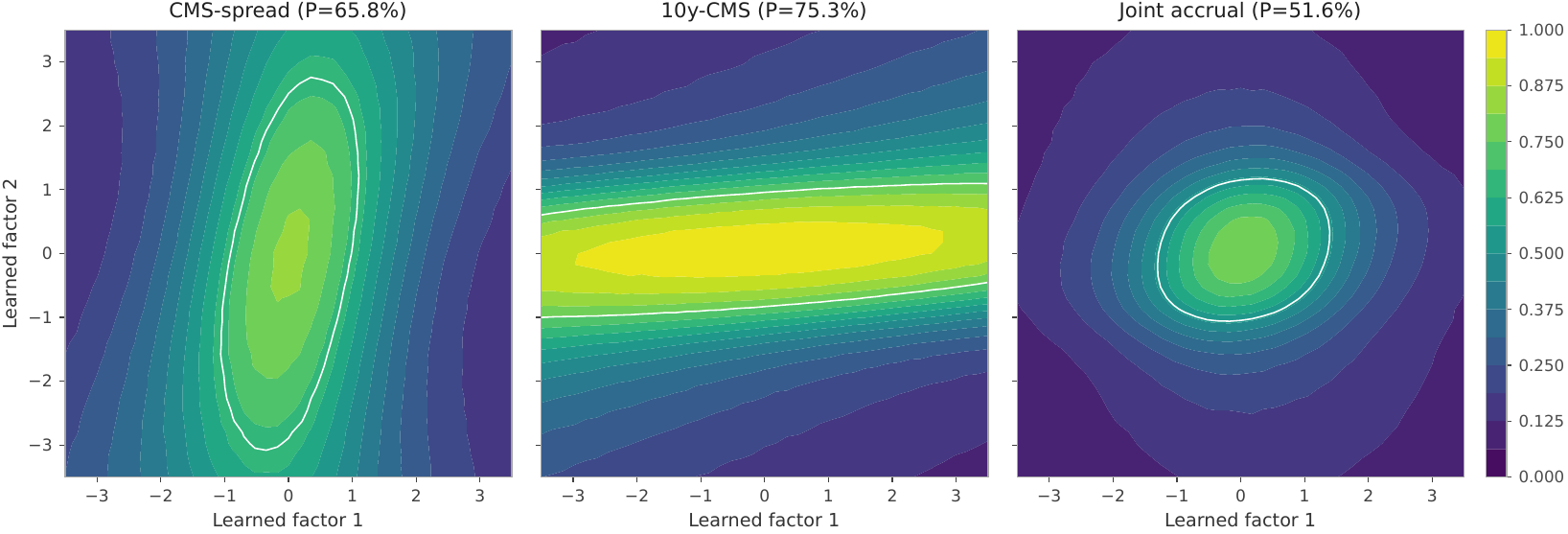}
\caption{Conditional expected fractions of monitoring dates satisfying the CMS spread condition, rate condition, or both.}
\label{fig:gauss2pp_conditional_structure}
\end{figure}

For the case $q=2$, write $Y=(Y_1,Y_2)^\top$ for the standardised learned factor.
\Cref{fig:gauss2pp_conditional_structure} reports, as functions of $y=(y_1,y_2)$, Monte Carlo estimates of the conditional expected fractions of monitoring dates on which the CMS-spread condition, the $10$-year CMS condition, and both conditions are satisfied. 
The white contours mark the estimated unconditional expected fractions shown in the panel titles.
We see that the two single-condition surfaces exhibit distinct directional structure: high conditional spread accrual is concentrated in a narrow tilted band, whereas high conditional $10$-year CMS accrual occupies a broader band with a markedly different orientation.
The different orientations indicate that the two accrual conditions depend on distinct payoff-relevant directions in the learned plane, which helps explain why $q=2$ improves on $q=1$ in \Cref{tab:gauss2pp-stratification}.

\subsection{Worst-of autocallable in the Wishart model}\label{subsec:worst-of-autocallable}
Our second example uses a Wishart stochastic-covariance model for three assets.
Under the pricing measure $\Prob$, let $W=(W^{ij})_{i,j=1}^3$ be a matrix Brownian motion whose entries are independent standard Brownian motions, and let $B$ be an independent standard Brownian motion in $\R^3$.
For the price vector $S=(S^1,S^2,S^3)^\top$, we set
\begin{equation*}
\begin{aligned}
\mathrm d\log S_t&=\left(r\mathbf 1-\delta-\frac12\operatorname{diag}(\Sigma_t)\right)\,\mathrm dt+\sqrt{\Sigma_t}\,\mathrm dZ_t,\\
\mathrm d\Sigma_t&=(\alpha Q^\top Q-\kappa\Sigma_t)\,\mathrm dt+\sqrt{\Sigma_t}\,\mathrm dW_tQ+Q^\top\mathrm dW_t^\top\sqrt{\Sigma_t},
\end{aligned}
\end{equation*}
with $\mathrm dZ_t=\mathrm dW_t\rho+\ell\,\mathrm dB_t$, $\ell\coloneqq\sqrt{1-\norm{\rho}^2}$, and $\Sigma_t$ representing the instantaneous covariance matrix of the log returns.
For our simulations, we take $\alpha=5$, $\kappa=1.5$, $r=0.02$, $S_0=100\, \mathbf 1$, $\delta=(0.015,0.020,0.025)^\top$, and $\rho=-(0.4/\sqrt3)\mathbf 1$.
Moreover, we set $\Sigma_0=\overline\Sigma$, where the long-run covariance matrix is given by
\begin{equation*}
\overline\Sigma=D
\begin{pmatrix}
1&0.65&0.45\\
0.65&1&0.50\\
0.45&0.50&1
\end{pmatrix}
D,
\qquad
D=\operatorname{diag}(0.20,0.23,0.26).
\end{equation*}
We choose $Q$ as the upper-triangular Cholesky factor satisfying $Q^\top Q=(\kappa/\alpha)\overline\Sigma$, so that the covariance drift becomes $\kappa(\overline\Sigma-\Sigma_t)$.
We consider a three-year horizon with uniform grid $t_k=k\Delta$, $k=0,\ldots,312$, with $\Delta=1/104$, and use a projected Euler scheme for $\Sigma$, replacing eigenvalues below $10^{-10}$ by $10^{-10}$ after each step.

For the payoff, we consider a unit-notional note with maturity $T=3$, a one-year non-call period, and quarterly observation dates $T_j=1+(j-1)/4$, $j=1,\ldots,9$.
At $T_j$, let $R_j^i\coloneqq S_{T_j}^i/S_0^i$ denote the relative performance of asset $i$ and set $R_j\coloneqq\min_{1\leq i\leq3}R_j^i$.
The autocall time is $\tau\coloneqq\min\{T_j:R_j\geq1\}$, where we stipulate that $\tau=\infty$ if the set $\{T_j:R_j\geq1\}$ is empty.
For an annual coupon rate of $c=0.08$, the discounted payoff is
\begin{equation*}
G\coloneqq
\begin{cases}
\exp(-r\tau)(1+c\tau), & \tau<\infty,\\[1mm]
\exp(-rT)(1+cT), & \tau=\infty\text{ and }R_9\geq0.60,\\[1mm]
\exp(-rT)R_9, & \tau=\infty\text{ and }R_9<0.60.
\end{cases}
\end{equation*}
Upon a call, the note pays the principal and the coupon accrued to the call date.
If no call occurs, the principal and the coupon accrued to maturity are paid when $R_9\geq0.60$, while redemption equals $R_9$ below the protection level.

As in the previous example, we consider a smooth proxy $\widetilde G$ to construct the derivative-based benchmarks.
Let $s_h(x)\coloneqq(1+\exp(-x/h))^{-1}$, with $h=0.02$, and set $\varepsilon=0.01$.
For $j=1,\ldots,9$, define $\pi_j\coloneqq s_h(\widetilde R_j-1)$, where
\begin{equation*}
\widetilde R_j\coloneqq-\varepsilon\log\left(\frac13\sum_{i=1}^3\exp\left(-\frac{R_j^i}{\varepsilon}\right)\right),
\quad
\omega_j\coloneqq\pi_j\prod_{k=1}^{j-1}(1-\pi_k),
\quad
\omega_\infty\coloneqq\prod_{k=1}^9(1-\pi_k),
\end{equation*}
and let $\pi_{\mathrm{prot}}\coloneqq s_h(\widetilde R_9-0.60)$.
The smooth payoff proxy is then defined as
\begin{equation*}
\widetilde G\coloneqq{}\sum_{j=1}^9\omega_j\exp(-rT_j)(1+cT_j)+\omega_\infty\exp(-rT)\left(\pi_{\mathrm{prot}}(1+cT)+(1-\pi_{\mathrm{prot}})\widetilde R_9\right).
\end{equation*}

To give some market context, as of September 2026, the products listed for trading on SIX Structured Products included, among others, three worst-of autocallable reverse convertibles from the Zürcher Kantonalbank group: a note with a 3.50\% annual coupon on the EURO STOXX~50, S\&P~500, and SMI indices maturing in August 2029 (ISIN CH1587222337), one with an 11.55\% annual coupon on Partners Group, Swissquote, and Temenos maturing in July 2028 (ISIN CH1573995839), and one with an 11.10\% annual coupon on Amrize, Avolta, and Logitech maturing in September 2027 (ISIN CH1560449014).
These securities share with our example the combination of a worst-of basket and an autocall feature, although our stylised contract is not intended to reproduce any one of them.

The long-run covariance matrix $\overline\Sigma$ motivates a model-based benchmark factor.
Let $\gamma_1\geq\gamma_2\geq\gamma_3>0$ denote the eigenvalues of $\overline\Sigma$, and let $e_1,e_2,e_3$ be corresponding orthonormal eigenvectors.
For $i=1,\ldots,3$, set
\begin{equation*}
Y_i^{\mathrm{PC}}\coloneqq\frac{1}{\sqrt{T\gamma_i}}\int_0^T e_i^\top\overline\Sigma^{1/2}\,\mathrm dZ_t,
\end{equation*}
which is the $i$th standardised principal component of the Gaussian vector $\overline\Sigma^{1/2}Z_T$.
For $q\in\{1,2,3\}$, the $q$-dimensional \emph{return-PC} factor is given by $(Y_1^{\mathrm{PC}},\ldots,Y_q^{\mathrm{PC}})^\top$.
Using the proxy $\widetilde G$, we also include the one-dimensional \emph{smooth-gradient} factor $\nabla\widetilde G(0)^\top\xi/\norm{\nabla\widetilde G(0)}$.

\begin{table}[tbp]
\centering
\begin{threeparttable}
\caption{Stratification results for a worst-of autocallable in the Wishart model.}
\label{tab:wishart-autocallable-stratification}
\begin{tabular}{@{}ccccccc@{}}
\toprule
q & Method & Bins & Estimate & SE & EV ratio & VR \\
\midrule
0 & crude MC & \textemdash{} & 0.94357532 & $4.208\times 10^{-4}$ & \textemdash{} & 1.000 \\
\midrule
1 & return-PC & 4096 & 0.94366679 & $2.894\times 10^{-4}$ & 0.524 & 2.115 \\
1 & smooth-gradient & 4096 & 0.94349270 & $3.655\times 10^{-4}$ & 0.243 & 1.326 \\
1 & GHS drift & 4096 & 0.94339664 & $3.696\times 10^{-4}$ & 0.246 & 1.296 \\
1 & GHS Hessian & 4096 & 0.94359345 & $3.699\times 10^{-4}$ & 0.233 & 1.294 \\
1 & \textbf{learned} & 4096 & 0.94353722 & $2.773\times 10^{-4}$ & \textbf{0.565} & \textbf{2.303} \\
\midrule
2 & return-PC & $64\times64$ & 0.94315629 & $2.671\times 10^{-4}$ & 0.590 & 2.482 \\
2 & GHS Hessian & $64\times64$ & 0.94334697 & $3.591\times 10^{-4}$ & 0.273 & 1.373 \\
2 & \textbf{learned} & $64\times64$ & 0.94301795 & $2.484\times 10^{-4}$ & \textbf{0.658} & \textbf{2.870} \\
\midrule
3 & return-PC & $16\times16\times16$ & 0.94310264 & $2.642\times 10^{-4}$ & 0.616 & 2.537 \\
3 & GHS Hessian & $16\times16\times16$ & 0.94339716 & $3.449\times 10^{-4}$ & 0.331 & 1.489 \\
3 & \textbf{learned} & $16\times16\times16$ & 0.94298518 & $2.494\times 10^{-4}$ & \textbf{0.695} & \textbf{2.847} \\
\bottomrule
\end{tabular}
\begin{tablenotes}[flushleft]
\footnotesize
\setlength{\leftmargin}{0pt}
\setlength{\itemindent}{0pt}
\setlength{\labelsep}{0pt}
\item[]$q$ denotes the factor dimension. SE: standard error; EV ratio: $\Var(\E[G\mid\sigma(Y)])/\Var(G)$; VR: variance reduction relative to crude Monte Carlo.
\end{tablenotes}
\end{threeparttable}
\end{table}

\Cref{tab:wishart-autocallable-stratification} shows differences across the competing factors.
Among the benchmarks, return-PC performs best in every dimension, with an estimated EV ratio increasing from about $52\%$ for $q=1$ to $62\%$ for $q=3$ and corresponding VR factors between $2.1$ and $2.6$.
This indicates that return variation already captures a substantial part of the low-dimensional structure relevant for the autocallable payoff, whereas the gradient- and GHS-based directions are less effective. 
Direct optimisation improves further on this benchmark: our learned factors attain EV ratios of about $57\%$, $66\%$, and $70\%$ for $q=1,2,3$, respectively, and give the largest VR factor for each dimension. 
Most of the realised gain is already obtained for $q=2$.
Although the EV ratio increases further for $q=3$, the VR factor decreases slightly from $2.870$ to $2.847$.

\begin{figure}[tbp]
\centering
\includegraphics[width=.9\linewidth]{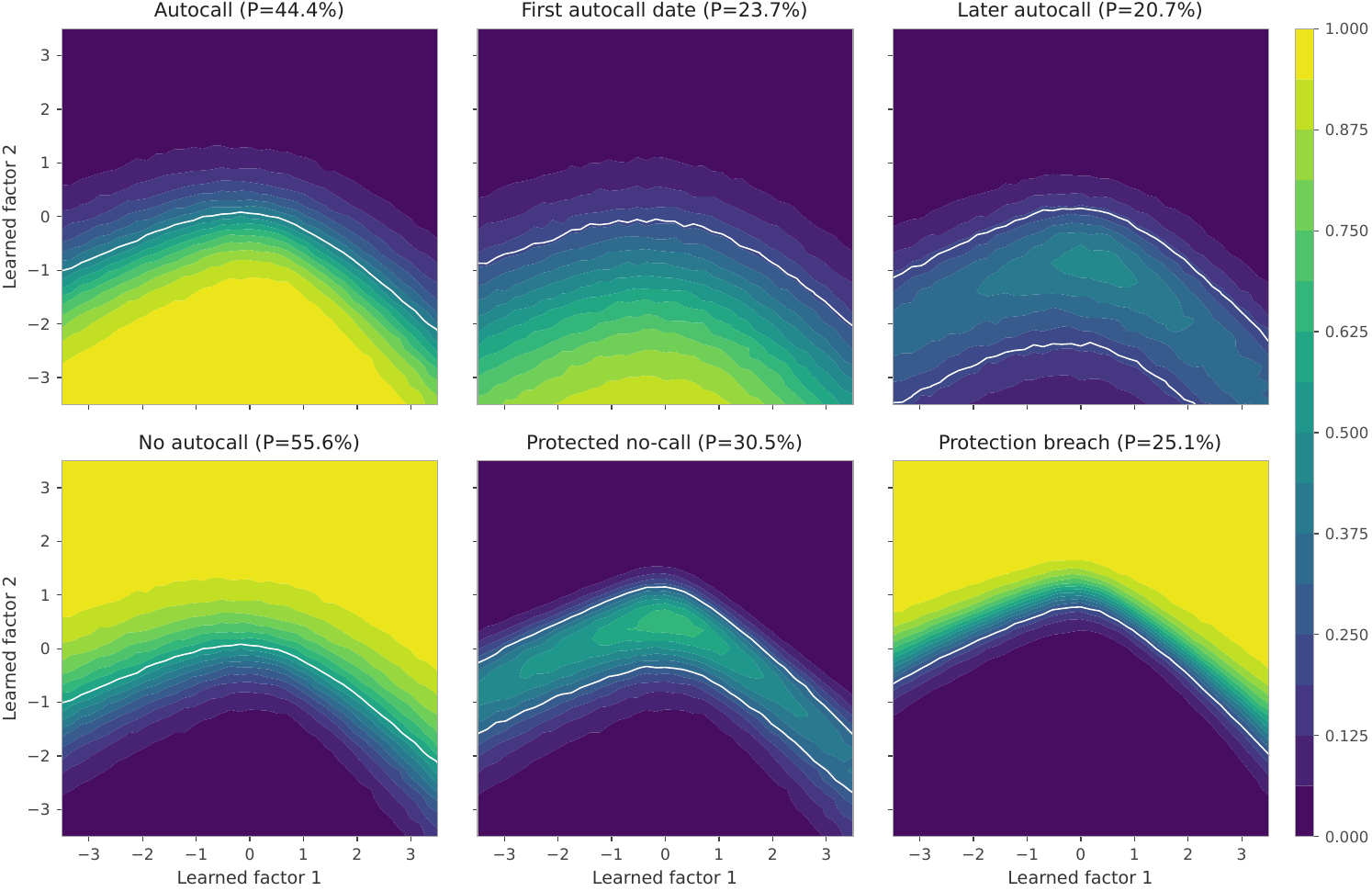}
\caption{Conditional probabilities of autocall and downside-protection events.}
\label{fig:wishart_autocallable_conditional_events}
\end{figure}

For $q=2$, the estimated conditional probabilities in \Cref{fig:wishart_autocallable_conditional_events} show how the learned factor separates the main contractual regimes.
One region is associated predominantly with redemption at the first eligible date, while an adjacent curved band shifts probability mass towards later redemption.
Beyond this transition, no-call outcomes become increasingly likely. 
Within the no-call region, a further transition separates protected no-call outcomes from breaches of the protection level.
What appears to be particularly interesting is that, as the factor realisation moves from the bottom to the top through the centre of the plane, the four regimes appear successively: conditional probability mass shifts from first-date redemption to later redemption, then to protected no-call outcomes, and finally to breaches of the protection level.
This suggests that the learned factor captures a structural transition across the different event types.

\section{Conclusion}\label{sec:conclusion}
In this paper, we studied the construction of stratification factors for derivative pricing under proportional allocation.
We established continuity results for conditional expectations and conditional distributions under approximation of Gaussian conditioning factors, and provided several examples illustrating the limits of this continuity theory.
Combined with density results for neural networks, we obtained a universal approximation theorem for admissible stratification factors that optimise the associated explained-variance criterion.
We illustrated the resulting approach for a constant-maturity-swap range accrual in the G2++ model and a worst-of autocallable in the Wishart model, where the learned factors compare favourably with several benchmark factors.
Our results provide a mathematical justification for approximating stratification factors for derivative pricing.

Our continuity results rely on a specific structure of Gaussian conditioning factors and eventual rank preservation of the factor covariance matrices.
The examples motivate a more precise characterisation of rank-changing limits for which continuity is retained.
A further extension is to consider nonlinear conditioning factors, for instance factors in a higher-order Wiener chaos.
Finally, our universal approximation result is nonconstructive, and future work could investigate the convergence of the numerical training procedure and the statistical properties of the learned factor and of the resulting stratified estimator.

\bibliographystyle{abbrvnat}
\bibliography{dstr-main}

\end{document}